\documentclass[]{jfm}

\usepackage{BOONDOX-cal,amssymb, amsmath,graphicx,color,rotating, 
natbib,booktabs}
\pdfoutput=1
\usepackage{newtxtext}
\usepackage{newtxmath}

\definecolor{webblue}{rgb}{0, 0, 0.5} 

\usepackage[pdfpagelabels,  
plainpages=true,
bookmarks=true,
bookmarksnumbered=true,
hypertexnames=true,
breaklinks=true,%
linkcolor=webblue,]{hyperref}
\hypersetup{
colorlinks  = true,
linkcolor = webblue,
urlcolor = webblue,
citecolor = webblue,
pdfauthor   = {John F. Rudge},
pdfsubject = {},
pdftitle    = {A crystallographic approach to symmetry-breaking in fluid layers},
pdfkeywords = {},
}

\renewcommand{\d}{\mathrm{d}}
\newcommand{\e}{\mathrm{e}}
\renewcommand{\i}{\mathrm{i}}
\newcommand {\dd}[2]{\frac{\d #1}{\d #2}}

\newcommand {\pdd}[2]{\frac{\partial #1}{\partial #2}}

\renewcommand{\v}{\boldsymbol{v}}
\newcommand{\zhat}{\hat{\boldsymbol{z}}}

\newcommand{\sig}{\mathsfi{\sigma}}
\newcommand\Ra{\mbox{\textit{Ra}}}

\begin{document}

\title{A crystallographic approach to symmetry-breaking\\ in fluid layers}
\author{John F. Rudge
  \corresp{\email{jfr23@cam.ac.uk}} and Dan McKenzie}

\affiliation{Bullard Laboratories, Department of Earth Sciences, 
University of Cambridge, Cambridge CB3 0EZ, UK}

\maketitle

\begin{abstract}
Symmetry-breaking bifurcations, where a flow state with a certain symmetry 
undergoes a transition to state with a different symmetry, are ubiquitous in 
fluid mechanics. Much can be understood 
about the nature of these transitions 
from symmetry alone, using the theory of groups and their representations. Here 
we show how the extensive databases on groups in crystallography can be 
exploited to yield insights into fluid-dynamical problems. In particular, we 
demonstrate the application of the crystallographic layer groups to problems in 
fluid layers, using thermal convection as an example. Crystallographic notation 
provides a concise and unambiguous description of the symmetries involved, and 
we advocate its broader use by the fluid dynamics community.
\end{abstract}

\section{Introduction}

One of the best known examples of pattern formation in fluid dynamics concerns 
Rayleigh-B\'{e}nard
convection in a fluid layer. As the temperature difference across the layer is 
increased the geometry of the flow changes from an initial stationary state in 
which no flow occurs, to a series 
of more complex flows. At the onset of convection patterns of rolls, hexagons, 
or squares can be seen, depending on the nature of fluid properties (e.g. 
whether the viscosity is temperature-dependent) and the nature of the boundary 
conditions. As the temperature difference is increased, further changes in the 
geometry of the flow occur, with the flow ultimately becoming chaotic and 
time-dependent.  

The transition from one flow geometry to another (e.g. from a stationary state 
to hexagons) involves a loss of symmetry; the system is said to have undergone 
a \emph{spontaneous symmetry-breaking bifurcation}. What is remarkable is that 
much can be 
understood about the nature of the bifurcation purely from the consideration of 
the symmetries of the system. This understanding comes from the subset of 
dynamical systems theory termed \textit{equivariant bifurcation theory}, and is 
well-covered in textbooks such as \citet{Hoyle2006, Golubitsky2002}. The 
language of 
symmetry is group theory. Each of the symmetry-breaking 
transitions from one state to another can be described by a state with a 
certain 
symmetry group transitioning to a state whose symmetry is a subgroup of the 
original group.

Crystallographers have long been concerned with transitions between states with 
different symmetry. Indeed, there is a celebrated theory of phase transitions 
in crystals due to Landau \citep{Landau1965}, which has much in common with 
equivariant 
bifurcation theory. Crystallographers have catalogued detailed symmetry 
information for periodic structures in the famous International Tables for 
Crystallography \citep{Hahn2006}, which have been supplemented in recent 
decades by extensive 
computer databases such as the Bilbao Crystallographic Server 
\citep{Aroyo2006b,Aroyo2006a}.

The aim of the present manuscript is demonstrate how the extensive databases on 
group theory in crystallography can be exploited to understand transitions in 
fluid layers. While there has already been extensive use of group theory 
to understand transitions in fluid layers, authors tend to use a bespoke 
notation for their particular problem. The advantage of crystallographic 
notation is that it is standardised. Moreover, there is a wealth of group-
theoretic information that can be simply looked up, without the need to be 
re-derived for each new problem. The use of crystallographic notation to 
describe convective transitions was first advocated by \citet{McKenzie1988}. 
The 
present manuscript is in a sense an extension of that work, and goes further by 
exploiting the theory of crystallographic layer groups \citep{Wood1964, 
Litvin1991}, which were only added to the International Tables in 2002 
\citep{Kopsky2010}.

There is no new theory discussed in this manuscript: the theoretical ideas are 
well-established and can be found in textbooks. We aim to 
provide here an informal introduction to the main ideas, and the interested 
reader can refer to the literature for the detailed theory. One of the main 
difficulties with this topic is the large amount of technical jargon needed to 
properly describe the ideas: the topic encompasses fluid dynamics, 
representation theory, bifurcation theory, and crystallography. Additional 
difficulties arise because different communities use different words for the 
same concept (e.g. factor group/ quotient group; invariant subgroup/ normal 
subgroup; isotropy group/ little group/ stabilizer). Where possible we have 
tried to use the notation of the International Tables for the crystallographic 
concepts, and the notation of the textbook by \citet{Hoyle2006} for equivariant 
bifurcation theory.    

The manuscript is organised as follows. In section \ref{sec:symlayer} we 
establish the fundamental symmetries of fluid layers. This is followed by an 
introduction to the crystallographic layer groups in section 
\ref{sec:layer_group} and an introduction to symmetry-breaking transitions in 
section \ref{sec:symbreak}. Section \ref{sec:reps} introduces the relevant 
representation theory, and section \ref{sec:eqbif} the relevant bifurcation 
theory. The theory is then applied to some simple convection problems in 
section 
\ref{sec:conv}. Three appendices provide additional technical details, and 
three supplements give tables of group theory information.

\section{The symmetry of fluid layers}\label{sec:symlayer}

We will consider a fluid dynamical problem which takes place in a layer. In 
terms of symmetry it is important to distinguish between three different 
symmetries: i) the symmetry of the domain; ii) the symmetry of the 
fluid-dynamical problem (i.e. 
the domain plus the governing equations and boundary conditions); and iii) the 
symmetry of solutions to the problem. Each of these symmetries may be different.

\subsection{Domain symmetries}

Let us consider first the symmetries of the domain. We have an infinite fluid 
layer, and will take $x$ and $y$ as horizontal co-ordinates, and $z$ as a 
vertical co-ordinate. Let $z=0$ denote the mid-plane of the layer, and $a$ 
denote the layer thickness. The domain is thus the region bounded by $-a/2 \leq 
z \leq a/2$.   

A symmetry of the domain is an invertible map which maps points in the domain 
to other points in the domain. Here we will consider only distance-preserving 
symmetries of the domain (isometries) as these will be the ones of relevance to 
the physical problem. We can translate all points by a horizontal displacement
vector $\boldsymbol{d} = (d_1, d_2, 0)$ 
\begin{equation}
 t_{\boldsymbol{d}}: (x, y, z) \rightarrow (x + d_1, y + d_2, z ) 
\label{eq:euc_1}
\end{equation}
and retain the same domain $-a/2 \leq 
z \leq a/2$. We also retain the same domain if we rotate about a vertical 
axis by an angle $\theta$,
\begin{equation}
 R^\theta_z: (x, y, z) \rightarrow (x \cos \theta - y \sin \theta, x \sin 
\theta + y \cos \theta, z ) \label{eq:euc_2}
\end{equation}
or reflect in a vertical mirror plane, e.g. with normal $x$,
\begin{equation}
 m_x: (x, y, z) \rightarrow ( - x, y, z). \label{eq:euc_3}
\end{equation}
The set of all such operations of the form \eqref{eq:euc_1}, \eqref{eq:euc_2}, 
\eqref{eq:euc_3} i.e. all horizontal translations, rotations about a vertical 
axes, and vertical mirrors, and their combinations, form a group known as 
$E(2)$, the Euclidean group 
of distance-preserving transformation in a plane. The fluid layer domain is 
also invariant 
under reflections in horizontal plane, i.e. with normal $z$,
\begin{equation}
 m_z: (x, y, z) \rightarrow (x, y, -z). \label{eq:euc_4}
\end{equation}
from which it also follows that the domain is also invariant under the 
inversion 
operation
\begin{equation}
 \overline{1}: (x, y, z) \rightarrow (-x, -y, -z). \label{eq:euc_5}
\end{equation}
The group of all distance-preserving operations (isometries) of the layer is 
$E(2) \times C_2$, a direct product of $E(2)$ and $C_2$, where $C_2$ denotes 
the cyclic group of 
order 2 containing two elements (taken here 
as the identity and the inversion 
operation). $C_2$ is sometimes denoted as $\mathbb{Z}_2$ in other work.
The combination of elements in the group $E(2) \times C_2$ leads to operations 
that are more 
complex than 
individual rotations and reflections e.g. one can have glide reflections which 
combine a reflection and a translation; and screw displacements which combine 
translations and rotations.

\subsection{Problem symmetries}

A fluid dynamical problem in the layer consists of the domain, a set of 
governing equations and boundary conditions. At each point in the domain there 
is a set of field variables that describe the state of the fluid (e.g. its 
temperature, velocity, pressure). A symmetry operation of the fluid 
dynamical problem is described by the combination of one of the isometries 
 with a description of how the field 
variables transform. If the governing equations and 
boundary conditions are invariant under this transformation then it is a 
symmetry of the fluid dynamical problem. 

Choices of material properties and boundary conditions mean that not all 
operations 
that are isometries of the domain are necessarily symmetries of the fluid 
dynamical 
problem. For example, if a different boundary condition is used on the top and 
bottom of the layer (e.g. fixed temperature on one, fixed-flux on another), 
then the system cannot be invariant under a horizontal mirror like 
\eqref{eq:euc_4}. Or, if one considers an inclined convection problem where 
the gravity vector is at an angle to the vertical axis of the layer then the 
problem is not invariant under arbitrary rotations about a vertical axis 
\citep{Reetz2020}.

A fluid dynamical problem which is invariant under the full group $E(2) \times 
C_2$ of isometries of the layer, and which will be used in many of the examples 
which follows, is Rayleigh-B\'{e}nard convection in a fluid layer of constant 
viscosity with appropriately chosen symmetric boundary conditions (e.g. both 
boundaries being fixed-flux and free-slip) and the Boussinesq approximation. 
The gravity vector is assumed to be aligned with the vertical. The natural 
field 
which describes the state of the system is the temperature. The governing 
equations are invariant under the operations given in  \eqref{eq:euc_1} to 
\eqref{eq:euc_5}, provided that the temperature perturbation $\theta$ (the 
difference in temperature from a conductive steady-state) transforms 
as
\begin{align}
 t_{\boldsymbol{d}}, R_z^\theta, m_x &: \theta \rightarrow \theta, \\
 m_z, \overline{1} &: \theta \rightarrow -\theta, \label{eq:mirror_th}
\end{align}
The sign change under horizontal mirror reflection is a manifestation of 
the symmetry between hot, rising fluid and cold, sinking fluid. A more detailed 
discussion of the symmetry of this problem can be found in Appendix 
\ref{sec:rb_sym}.

\subsection{Solution symmetries}

In general, the symmetries of solutions to the equations are not the same as 
symmetries of the problem, although the solutions' symmetries are generically 
subgroups 
of the set of symmetries of the problem. Rayleigh-B\'{e}nard convection 
provides a natural example of this: a planform of hexagons or squares is 
not invariant under any arbitrary translation but only a subgroup of allowed 
translations. However, one can apply a general element of the symmetry group of 
the problem to a given solution to yield another solution of the equations.

\begin{figure}
\centering
\includegraphics[width=\columnwidth]{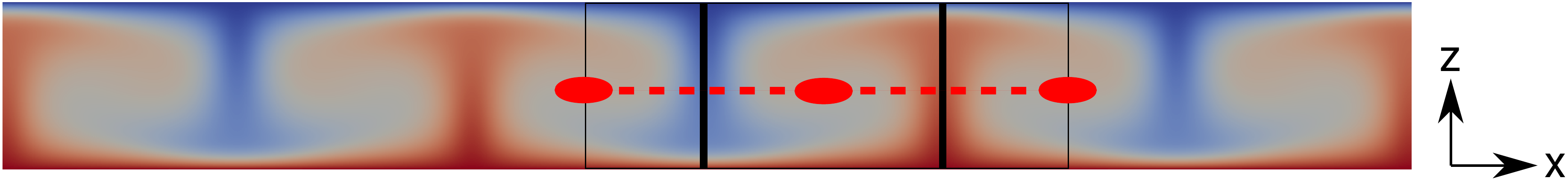}
\caption{An illustration of the symmetries of a simple convective flow. Shown 
is a temperature field from a 2-dimensional numerical simulation of 
constant-viscosity Rayleigh-B\'{e}nard convection with free-slip, 
fixed-temperature boundary conditions. This can also be considered as 
a temperature field in 3-dimensions for convective rolls if the field is 
continued into the page (the $y$-direction). The Rayleigh number 
is $10^4$ and the flow is steady. The flow pattern is periodic in the 
$x$-direction and the repeating unit cell is identified by the thin black 
lines. Thick black lines indicate vertical mirror planes ($m_x$). Red ovals 
indicate 
two-fold horizontal rotation axes ($2_y$). The red colouring of the oval is 
used to 
indicate the symmetry involves a change of sign of the temperature field (i.e. 
changing from hot upwelling in red to cold downwelling in blue). The dashed-red 
horizontal line 
indicates a horizontal glide plane: the pattern is invariant after translating 
in the $x$-direction by half the width of the unit cell, reflecting in the 
horizontal mid-plane and changing the sign of the temperature field. The rolls 
in 3-dimensions also have a continuous translation symmetry in the 
$y$-direction 
and $m_y$ mirrors.}
\label{fig:2dexample}
\end{figure}

\autoref{fig:2dexample} illustrates the symmetries of a particular solution to 
Rayleigh-B\'{e}nard convection, namely that of convective rolls. Unlike the 
governing equations which have a continuous translation symmetry in the 
$x$-direction, the rolls have a discrete periodicity. Also unlike the 
governing equations, the solution is not invariant under a horizontal mirror 
$m_z$, but it is invariant when the horizontal mirror $m_z$ is combined with a 
translation by half the width of the unit cell: this is an example of a glide 
reflection. 

\section{Crystallographic layer groups}\label{sec:layer_group} 

This work focuses on particular subgroups of $E(2) \times C_2$ known as the 
crystallographic layer groups. They are the set of isometries of the layer that 
are doubly-periodic in space: that is, instead of having continuous translation 
symmetry in the horizontal as $E(2) \times C_2$ does, the translation symmetry 
is discrete. The layer groups are invariant under $t_{\boldsymbol{d}}$ in 
\eqref{eq:euc_1} only for discrete lattice vectors satisfying
\begin{equation}
 \boldsymbol{d} = x \boldsymbol{a}_1 + y \boldsymbol{a}_2
\end{equation}
where $x, y \in \mathbb{Z}$, and $\boldsymbol{a}_1$, $\boldsymbol{a}_2$ are 
basis 
vectors for the lattice. Many pattern-forming problems lead to steady fluid 
flows that can be described as having a layer group symmetry. For example 
planforms described as squares, hexagons, bimodal, triangles, are all 
doubly periodic in space and are examples of layer group symmetry. The 
principal example of a convective flow that is not a layer group symmetry is 
that of convective rolls: this has a discrete translation symmetry in one 
horizontal direction, but a continuous translation symmetry in another 
horizontal direction. Layer groups are an example of a subperiodic group: a 
group where the dimension of the space is greater than the dimension of the 
periodic lattice. For layer groups, the space in which the group elements act 
is 3-dimensional, but there is only a 2-dimensional lattice of translations. 
The layer groups are in a sense intermediate between full 3-d space groups 
(3-dimensional groups with a 3-dimensional translation lattice) and the 2-d 
plane or wallpaper groups (2-dimensional groups with a 2-dimensional 
translation lattice).

\begin{figure}
\centering
 \includegraphics[width = 0.7\columnwidth]{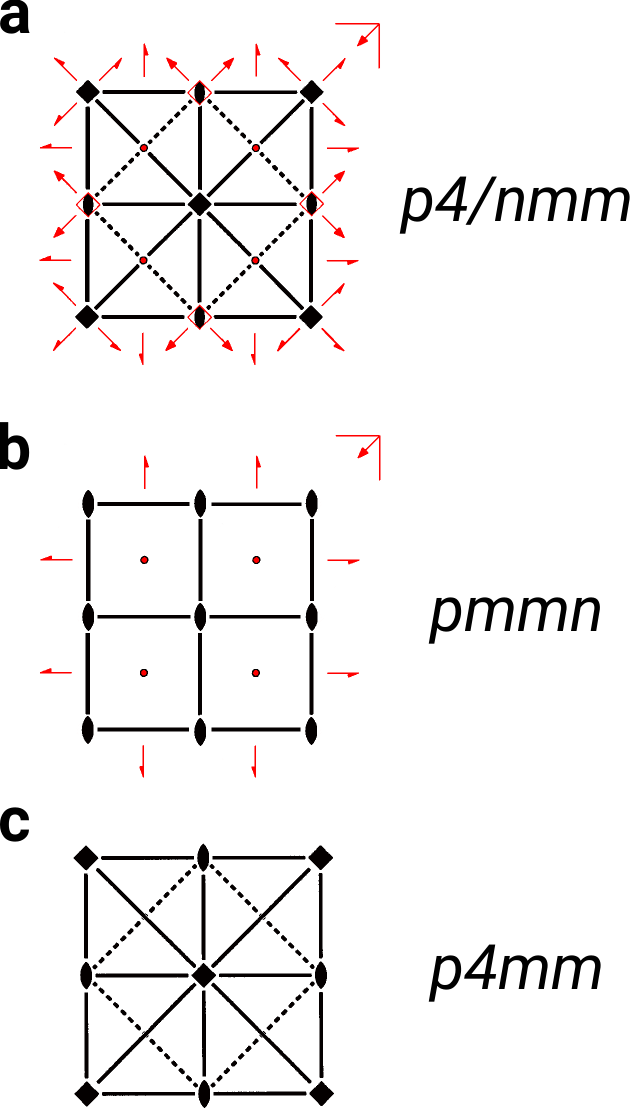}
 \caption{Symmetry diagrams from ITE for a) $p4/nmm$ (origin choice 1), and two 
of its subgroups, b) $pmmn$ and c) $p4mm$. Shown is the
unit cell in a projection onto the horizontal mid-plane.
 Squares 
indicate fourfold vertical rotation axes ($4_z$), filled ovals are twofold 
vertical rotation axes ($2_z$), circles are inversion centres 
($\overline{1}$), unfilled squares with filled ovals indicate a 
$\overline{4}_z$ 
vertical inversion axis ($\overline{4}_z$ combines the fourfold 
vertical rotation $4_z$ with the inversion operation $\overline{1}$). Solid 
lines are vertical mirror planes, dashed lines 
are vertical glide planes. The symbol in the top right refers to the horizontal 
glide plane $n$, where the symmetry operation combines a vertical mirror $m_z$ 
with a translation by $(\tfrac{1}{2}, \tfrac{1}{2}, 0)$. Full arrows around the 
edge refer to a horizontal twofold 
rotation axis, half-arrows refer to a twofold screw axis. Red 
colouring 
indicates symmetry operations that send $z \rightarrow -z$ and will be 
associated with sign changes in the temperature field (hot to cold and 
vice-versa). Examples of 
convective flows with these symmetries are shown in \autoref{fig:examples1}b; 
\autoref{fig:examples2}b;
\autoref{fig:convection_break}, \autoref{fig:3drenders} and 
\autoref{fig:shift}.}
\label{fig:p4nmm_sym}
\end{figure}

There are 80 layer groups, and their properties are detailed in the 
International Tables for Crystallography, volume E \citep{Kopsky2010} 
(hereafter referred to as ITE) and in computer databases 
such as the Bilbao Crystallographic Server \citep{DelaFlor2021} (hereafter 
referred to as BCS). Each layer group is identified by a unique 
number and Hermann-Mauguin symbol. One example that we will focus on is the 
layer group $p4/nmm$ (layer group 64, illustrated in \autoref{fig:p4nmm_sym}a). 
This group has a square lattice, a 4-fold vertical rotation axis, two conjugate 
sets of vertical mirror planes, and a glide reflection $n$ that combines 
reflection in a horizontal plane with a translation by $(\tfrac{1}{2}, 
\tfrac{1}{2}, 0)$. The first letter of the Hermann-Mauguin symbol denotes the 
centering type of the conventional unit cell: for layer groups this is either 
$p$ for primitive cell or $c$ for centered cell. The next one to three elements 
of the symbol describe symmetry elements about different axes. For $p4/nmm$ the 
three elements are $4/n$, $m$ and $m$. The $4/n$ element denotes the presence 
of both the 4-fold vertical rotation axis and the glide reflection $n$, 
whose glide plane also has a vertical normal (the slash indicates that the 
rotational symmetry axis and the normal to the glide plane are parallel). The 
next two elements labelled $m$ represent vertical mirror planes. Other 
possible Hermann-Mauguin elements found in other layer group symbols include 
the 
symbol $\overline{1}$ representing the inversion operation of \eqref{eq:euc_5}, 
the symbols $\overline{3}$, $\overline{4}$, $\overline{6}$ which combine 
three-, four-, or six-fold rotation with an inversion, and $a$ or $b$ for glide 
reflections with translations parallel to the basis vectors of the lattice 
$\mathbf{a}_1$ and $\mathbf{a}_2$ respectively.

From \eqref{eq:mirror_th} we have that a symmetry operation that sends $z 
\rightarrow - z$ involves a change in sign of the temperature perturbation 
(i.e. hot to cold or vice-versa). There is a broader class of crystallographic 
groups termed ``black and white'', ``magnetic'' or ``Shubnikov'' that have as a 
possible group element $1^\prime$ which changes the sign of a field without 
changing position. With such groups a combination of a horizontal mirror and a 
sign change would be denoted as $m_z^\prime$, and the black-and-white layer 
groups depicted in \autoref{fig:p4nmm_sym}a,b  would be referred to as 
$p4/n^\prime 
m m$ and $pmmn^\prime$ \citep{Litvin2013}. 
However, here we will not denote the symmetry operations with primes for two 
reasons: First, the fluid problems we consider are 
invariant only on combining the sign change in $\theta$ with the horizontal 
mirror: they are not invariant under a sign change in $\theta$ alone, so the 
$1^\prime$ operator is not present. Second, a general fluid dynamical problem 
can consist of more field variables that just one, and each variable 
may transform in a different way under the isometries e.g. the horizontal 
velocities and the toroidal potential do not change sign under $m_z$ (see 
\autoref{sec:rb_sym}). We will simply write $m_z$ as the group element 
corresponding to horizontal mirror reflection and it should be understood that 
it acts on different fields in different ways (some change sign, some do not).

Many of the plots in this manuscript show the temperature field in the 
horizontal mid-plane. Position in the mid-plane is 
invariant under the horizontal mirror $m_z$: the only action of $m_z$ in the 
mid-plane is to change the sign of the temperature perturbation. The 
mid-plane temperature fields can therefore be considered as directly belonging 
to one of 
the two-dimensional black-and-white plane groups. There are 80 black-and-white 
plane groups, which are isomorphic to the 80 layer groups. A mapping between 
the symbols used for black-and-white plane groups and those for layer groups 
can 
be found in ITE.

\section{Symmetry-breaking transitions}\label{sec:symbreak}

Suppose that as a control parameter (such as the Rayleigh number) is varied, a 
symmetry-breaking transition occurs from a state with symmetry group $G$ to a 
state with a lower-symmetry group $H$. For simplicity, let us just consider 
steady states. Purely from symmetry arguments alone  
there is often much that can be said about the nature of the transition: e.g. 
one can often classify the nature of the bifurcation as being either pitchfork 
or transcritical, and also write down the generic form of the equations 
describing the amplitudes of the critical modes (section \ref{sec:eqbif}). More 
generally, given an 
initial state with a group $G$ one can determine the possible groups $H$ that 
can arise in a symmetry-breaking transition.

The first requirement for $H$ is that it is a subgroup of $G$. Subgroups of 
layer groups have a particular structure and classification \citep{Muller2013} 
where the letters t or k refer to the symmetries that are retained. 
A subgroup is 
termed a \emph{translationengleiche} subgroup or t-subgroup if it has the same 
translation symmetries as its parent. A t-subgroup has a different point group 
to its parent and so is necessarily a non-isomorphic subgroup with a different 
layer group number and symbol. A subgroup is termed a \emph{klassengleiche} 
subgroup or k-subgroup if the translations are reduced but the order of the 
point group remains the same. Klassengleiche subgroups can be further 
categorised into those that are isotypic (have the same layer group number and 
symbol) and those that are non-isotypic (have a different layer group symbol). 
Finally a subgroup may have both the order of the point group reduced and the 
translations reduced. However in this case there always exists an intermediate 
subgroup $M$ such that $M$ is a t-subgroup of $G$ and $H$ is a k-subgroup of 
$M$ (Hermann's theorem). As such any subgroup $H$ of $G$ can be described in 
terms of a chain of t and k relationships. 

\begin{table}
\centering
\caption{Maximal subgroups of $p4/nmm$ (layer group no. 
64)}\label{tab:maxsub_p4nmm}
\begin{tabular}{ccccccccc}
\toprule
Subgroup & HM symbol & Index & Type & Factor group & Core & Core HM & Image & 
Bifurcation \\
\midrule
46 & $pmmn$ & 2 & t & $C_2$ & 46 & $pmmn$ & $C_1$ & pitchfork \\
48 & $cmme$ & 2 & t & $C_2$ & 48 & $cmme$ & $C_1$ & pitchfork \\
52 & $p4/n$ & 2 & t & $C_2$ & 52 & $p4/n$ & $C_1$ & pitchfork \\
54 & $p42_12$ & 2 & t & $C_2$ & 54 & $p42_12$ & $C_1$ & pitchfork \\
55 & $p4mm$ & 2 & t & $C_2$ & 55 & $p4mm$ & $C_1$ & pitchfork \\
58 & $p\overline{4}2_1m$ & 2 & t & $C_2$ & 58 & $p\overline{4}2_1m$ & $C_1$ & 
pitchfork \\
59 & $p\overline{4}m2$ & 2 & t & $C_2$ & 59 & $p\overline{4}m2$ & $C_1$ & 
pitchfork \\
64 & $p4/nmm$ & 9 & k & $C_3^2 \rtimes D_4$ & 5 & $p11a$ & $D_4$ & 
transcritical 
\\
64 & $p4/nmm$ & 25 & k & $C_5^2 \rtimes D_4$ & 5 & $p11a$ & $D_4$ & pitchfork \\
64 & $p4/nmm$ & 49 & k & $C_7^2 \rtimes D_4$ & 5 & $p11a$ & $D_4$ & pitchfork \\
\bottomrule
\end{tabular}
\end{table}

A subgroup $H$ of a group $G$ is termed \emph{maximal} if there is no 
intermediate subgroup $M$ of $G$ such that $H$ is a proper subgroup of $M$. 
Both ITE and BCS provide comprehensive lists of the maximal subgroups of the 
layer groups, from which parent group-subgroup relationships can be 
described. An example of such subgroup information is given in 
\autoref{tab:maxsub_p4nmm} for the layer group $p4/nmm$, along with additional 
information useful for describing symmetry-breaking bifurcations. Similar 
tables for all 80 layer groups can be found in supplement 1. In 
\autoref{tab:maxsub_p4nmm} 
each maximal subgroup is listed with its layer group number and Hermann-Mauguin 
symbol. The index of the subgroup in the parent is given (the index is the 
number of left cosets of the subgroup in the parent). The type of subgroup is 
given as either t or k for translationengleiche or klassengleiche. The 
nature of the additional information in \autoref{tab:maxsub_p4nmm} on factor 
group, core, and 
bifurcation is described in sections \ref{sec:reps} and \ref{sec:eqbif}.  

Group theory places more constraints on the group $H$ than it simply 
being a subgroup of $G$. In fact for a generic steady-state symmetry-breaking 
bifurcation the subgroup $H$ must be an \textit{isotropy subgroup} of a 
particular \textit{absolutely irreducible representation} of the group $G$ 
\citep{Hoyle2006,Golubitsky2002}. 
Thus to understand symmetry-breaking bifurcations of layer groups we must 
understand their group representations, which we turn to now. 

\section{Representations of layer groups}\label{sec:reps}

A \emph{representation} of a group is simply a mapping of the group elements to 
a set of matrices in a way that preserves the group operation (i.e. the mapping 
is a homomorphism onto $GL(V)$). A representation acts on a certain vector 
space 
$V$ of dimension $n$. An \emph{invariant subspace} of a representation is a 
vector subspace $W$ that has the property that $\mathsfbi{D} \boldsymbol{w} \in 
W$ 
for all $\boldsymbol{w} \in W$ and for all $\mathsfbi{D}$ in the set of 
representation 
matrices. The 
spaces 
$W=\lbrace\boldsymbol{0} \rbrace$ and $W=V$ are always invariant subspaces, 
known 
as the trivial subspaces. If the representation contains a non-trivial 
invariant 
subspace then it is said to be \emph{reducible}, otherwise it is 
\emph{irreducible}. A representation is 
\textit{absolutely irreducible} if the only linear maps which commute with the 
representation are multiples of the identity. For representations over 
$\mathbb{C}$ there is no distinction between being absolutely irreducible and 
just irreducible, but there is a difference over $\mathbb{R}$ where 
representations can be irreducible but not absolutely irreducible. Irreducible 
representations (or irreps for short) are the building blocks of representation 
theory. Any representation of the group can be written as a direct sum of its 
irreducible representations (Maschke's theorem). 

Given a point $\boldsymbol{v} \in V$, we can define its \emph{isotropy 
subgroup} 
$\Sigma$ as
\begin{equation}
\Sigma= \lbrace { g \in G : g \boldsymbol{v}  = 
\boldsymbol{v}} \rbrace
\end{equation}
and its corresponding \textit{fixed-point subspace} by
\begin{equation}
\operatorname{Fix} (\Sigma) = \lbrace {\boldsymbol{w} \in V :  g 
\boldsymbol{w}  = 
\boldsymbol{w}, \forall g \in \Sigma} \rbrace
\end{equation}
An isotropy subgroup is said to be \emph{axial} if the dimension of its 
fixed-point subspace is 1. Axial isotropy subgroups are of particular interest 
because the existence of solution branches with the given isotropy subgroup is 
guaranteed under certain conditions by the equivariant branching lemma 
\citep{Hoyle2006,Golubitsky2002}.

Classifying the steady-state symmetry-breaking bifurcations of layer groups 
consists of identifying their irreducible representations, and 
subsequently finding their isotropy subgroups. The general theory of 
irreducible 
representations of layer groups is in detail somewhat involved, but it is 
known, and results can simply be looked up in textbooks or extracted from 
computer 
databases. In many cases it is not necessary to invoke the full general theory, 
as the appropriate irreps can be found quickly by lifting from an appropriate 
factor group. 

\subsection{Lifting representations}\label{sec:lift}

Given a group $G$ and a normal subgroup $N$, one can form the factor group (or 
quotient group) $G/N$. The elements of $G/N$ are the left cosets of $N$ in $G$, 
which have a well-defined multiplication operator when the subgroup $N$ is 
normal.

Suppose we are interested in understanding a transition between a group $G$ and 
a subgroup $H$. We want to know the irrep of $G$ associated with the 
transition. In the crystallography literature this is termed ``the 
inverse Landau problem'' \citep{Ascher1977, Litvin1986}. One solution to this 
is 
as follows. We first find the 
normal core $N$ of the subgroup $H$ in $G$: that is, the largest normal 
subgroup of $G$ that is contained in $H$. In some cases this may be the whole 
subgroup $H$, but not in general. We then form the factor group $G/N$, and we 
refer to this as the factor group associated with the transition. 
\autoref{tab:maxsub_p4nmm} gives the factor groups and normal cores associated 
with each of the maximal subgroups of $p4/nmm$. The table also gives the image 
of the subgroup $H$ under the natural homomorphism onto cosets of $N$. The 
advantage of finding the factor group $G/N$ is that it is typically a small 
finite group and so finding its irreps is much more straightforward than the 
finding the irreps for the group $G$ (which in the case of layer groups is an 
infinite group). Moreover, the irreps of the 
factor group $G/N$ can be \emph{lifted} to an irrep of the group $G$ using the 
natural homomorphism onto cosets. Suppose we have an irrep $\rho$ of the factor 
group
\begin{equation}
 \rho: G/N \rightarrow GL(V)
\end{equation}
and suppose $q$ is the natural homomorphism
\begin{equation}
 q : G \rightarrow G/N,
\end{equation}
where $q(g) = g N$. Then the composition $\rho \circ q$ is the irrep of $G$ 
lifted from $G/N$. Indeed it is an irrep of $G$ with $N$ in its kernel. 
Representations lifted from a factor group are sometimes termed 
\emph{engendered representations} in the crystallography literature. 

All the t-subgroups of $p4/nmm$ have the same factor group: $C_2$. In this case 
the subgroups are all normal subgroups, so the normal core $N$ is the same as 
the 
subgroup $H$. The irreps associated with these transitions 
are
very simple. They are 1-dimensional and just send each element of the subgroup 
$H$ to 1 
and all others to $-1$. As will be discussed later, this is associated with a 
pitchfork bifurcation. Any index-2 subgroup necessarily has a factor group of 
$C_2$ and is associated with a pitchfork bifurcation.  

\subsection{t-subgroups}

The irreps associated with translationengleiche transitions, which preserve the 
translations of the lattice, can be found by lifting from an appropriate 
factor group. The set of all translations $T$ of the lattice forms a normal 
subgroup of any layer group. Therefore the irreps of $G$ with the 
pure translations in the kernel can be found by lifting from the factor group 
$G/T$. The factor group $G/T$ is isomorphic to the isogonal point group 
associated with the layer group, so the irreps are simply those of the 
corresponding point group.   

\begin{table}
\centering

\begin{tabular}{r|rrrrrrrrrr|l}
\toprule
 & $1$ & $2_z$ & $4_z$ & $2_y$ & $2_{xy}$ & $\overline{1}$ & $m_z$ & 
$\overline{4}_z$ & $m_y$ & $m_{xy}$ & axial subgroups \\
\midrule
size & 1 & 1 & 2 & 2 & 2 & 1 & 1 & 2 & 2 & 2 &  \\ \midrule
$A_{1g}$ & $1$ & $1$ & $1$ & $1$ & $1$ & $1$ & $1$ & $1$ & $1$ & $1$ & $p4/nmm$ 
(64) \\
$A_{2g}$ & $1$ & $1$ & $1$ & $-1$ & $-1$ & $1$ & $1$ & $1$ & $-1$ & $-1$ & 
$p4/n$ (52) \\
$B_{1g}$ & $1$ & $1$ & $-1$ & $1$ & $-1$ & $1$ & $1$ & $-1$ & $1$ & $-1$ & 
$pmmn$ (46) \\
$B_{2g}$ & $1$ & $1$ & $-1$ & $-1$ & $1$ & $1$ & $1$ & $-1$ & $-1$ & $1$ & 
$cmme$ (48) \\
$E_{g}$ & $2$ & $-2$ & $0$ & $0$ & $0$ & $2$ & $-2$ & $0$ & $0$ & $0$ & 
$p2_1/m11$ (15), $c2/m11$ (18) \\
$A_{1u}$ & $1$ & $1$ & $1$ & $1$ & $1$ & $-1$ & $-1$ & $-1$ & $-1$ & $-1$ & 
$p42_12$ (54) \\
$A_{2u}$ & $1$ & $1$ & $1$ & $-1$ & $-1$ & $-1$ & $-1$ & $-1$ & $1$ & $1$ & 
$p4mm$ (55) \\
$B_{1u}$ & $1$ & $1$ & $-1$ & $1$ & $-1$ & $-1$ & $-1$ & $1$ & $-1$ & $1$ & 
$p\overline{4}2_1m$ (58) \\
$B_{2u}$ & $1$ & $1$ & $-1$ & $-1$ & $1$ & $-1$ & $-1$ & $1$ & $1$ & $-1$ & 
$p\overline{4}m2$ (59) \\
$E_{u}$ & $2$ & $-2$ & $0$ & $0$ & $0$ & $-2$ & $2$ & $0$ & $0$ & $0$ & 
$pm2_1n$ 
(32), $cm2e$ (36) \\
\bottomrule
\end{tabular}
\caption{Translationengleiche character table of $p4/nmm$ (No. 
64). The top row gives the Seitz symbol labels for a member of each conjugacy 
class. The number of elements in each conjugacy class is listed on the row 
beneath. Each irrep is given a label on the left using Mulliken 
notation. The right column gives the corresponding axial isotropy subgroups 
associated with each irrep. Note that the Seitz symbol labels only 
refer to the point group part of the symmetry operations: the coset 
representatives of  
$2_y$, $2_{xy}$, $\overline{1}$, $m_z$, $\overline{4}_z$ also involve a 
translation by $\left(\tfrac{1}{2}, \tfrac{1}{2}, 0\right)$ (see the ITE 
description of 
$p4/nmm$, origin choice 1).}\label{tab:charp4nmm}

\end{table}

The character table for the factor group $G/T$ is shown for $p4/nmm$ in  
\autoref{tab:charp4nmm}, and is the same as that for isogonal point group 
$4/mmm$ ($D_{4h}$). Similar 
tables for all 80 layer groups can be found in supplement 2. The 
\emph{character} of a representation matrix is 
simply its trace. Characters are independent of the basis used in the 
representation, and are the same for group elements that are conjugate. A 
character table simply consists of a table of all the characters for all the 
irreps of a group. For 
many applications of representation theory it is sufficient to know the 
characters 
of the representation and it is not necessary to know the representation 
matrices themselves. 

Each of the 7 t-subgroups listed in \autoref{tab:maxsub_p4nmm} is associated 
with one of the 1-dimensional irreps in \autoref{tab:charp4nmm}. There are also 
additional axial subgroups identified in \autoref{tab:charp4nmm} associated 
with the 2-dimensional representations labelled $E_g$ and $E_u$. These 
subgroups are not in \autoref{tab:maxsub_p4nmm} as they are not maximal 
subgroups. It should be stressed that isotropy subgroups need not be maximal 
subgroups.

\subsection{General theory of representations of layer groups}

Irreps associated with k-transitions, where translation symmetries are lost, 
can 
also be obtained by lifting from appropriate factor groups. An example is given 
in \autoref{tab:maxsub_p4nmm} which lists an index-9 k-transition from 
$p4/nmm$ to $p4/nmm$ where the associated irreps could be found by 
considering the irreps of the corresponding factor group $C_3^2 \rtimes D_4$ 
(in this symbol $C_3^2$ denotes the direct product of $C_3$ with itself, i.e. 
$C_3 \times C_3$; the symbol $\rtimes$ denotes the semi-direct product). A 
discussion of the irreps of this particular factor group can be found in 
\citet{Matthews2004} (see his Figure 1). Such a transition is an example of a 
spatial-period-multiplying bifurcation where the periodicity of the pattern is 
broken but maintained on a larger scale: in this case after the symmetry break 
the lattice basis vectors are scaled by a factor of 3 in each direction.  

An alternative approach is to exploit the general 
theory which describes the complete set of irreps 
of layer groups. This theory is somewhat involved, but is understood, and one 
can simply look up appropriate representations using published tables 
\citep{Bradley1972, Litvin1991, Milosevic1998} or computer software 
\citep{Aroyo2006b, DelaFlor2021, Stokes2016}.

The starting point for the general theory concerns the representations of the 
subgroup $T$ of all translations of the lattice. The subgroup $T$ is a normal 
subgroup of any layer group. It is also an Abelian group, so its 
irreps are 1-dimensional. The irreps of $T$ are simply
$\e^{-\i \boldsymbol{k} \bcdot \boldsymbol{d}}$ for a translation by 
a vector 
$\boldsymbol{d}$, where $\boldsymbol{k}$ is a wavevector which labels the 
particular 
irrep. Wavevectors that differ by a reciprocal lattice vector lead to identical 
irreps. As such, the wavevectors for defining irreps are restricted to a 
region of reciprocal space known as the Brillouin zone (a unit cell in 
reciprocal space) such that each 
irrep has a unique $\boldsymbol{k}$ label.

The irreps of the layer groups can built up from the irreps of $T$ using the 
theory of induced representations (see appendix \ref{sec:rep_nonzero} for a 
brief example, and \citet{Bradley1972,Aroyo2006b,DelaFlor2021} for the detailed 
theory). Each irrep is labelled by a wavevector $\boldsymbol{k}$, a symbol 
which 
represents the type of wavevector (the labels $\Gamma$, $\Sigma$, $\Delta$ etc 
in \autoref{fig:brillouin} of appendix \ref{sec:rep_nonzero}), and an index (1, 
2, 3, ...) referencing a 
particular  representation of the little group of the wavevector. For example, 
the index-9 k-transition from 
$p4/nmm$ to $p4/nmm$ is associated with two possible irreps of the parent 
group: $^* \Sigma_1$ with 
$\boldsymbol{k} = (1/3, 1/3)$ and $^* \Delta_3$ with  
$\boldsymbol{k} = (0, 1/3)$. The full matrices of these representations is 
given in 
appendix \ref{sec:p4nmm_reps}. The irreps associated with t-transitions have a 
zero wavevector, $\boldsymbol{k}=(0, 0)$. As such are sometimes labelled by 
$^*\Gamma$ and an index, rather than the Mulliken symbols used in 
\autoref{tab:charp4nmm}, as they correspond to the $\Gamma$ point in the 
Brillouin zone (\autoref{fig:brillouin}). 

Much information can be obtained about the irreps and isotropy subgroups 
associated with transitions by querying computer databases 
\citep{DelaFlor2021,Aroyo2006b,Stokes2016,Perez-Mato2012,Iraola2022}. Given a 
parent group and a subgroup one can ask the software tools to provide the 
associated irreps and the corresponding fixed-point subspaces of the 
isotropy subgroups. For a given parent group one can also obtain from the tools 
a 
complete listing of all possible isotropy subgroups and the corresponding 
irreps. Most of these software tools are designed for use on full 3D space 
groups, rather than layer groups. However, each layer group can be associated 
with a corresponding space group \citep{Litvin2000}. Given a 3D space group 
$S$, and $T_z$ as the 
one-dimensional subgroup of $S$ of the vertical translations, then the factor 
group $S / T_z$ is isomorphic to a layer group. The irreps of layer groups can 
be obtained from the irreps of space groups where the wavevector is constrained 
to lie in a particular plane. 

\section{Equivariant bifurcation theory}\label{sec:eqbif}

Once the irrep associated with a particular transition is known, then the 
nature 
of the bifurcation can be understood using equivariant bifurcation theory 
\citep{Hoyle2006,Crawford1991,Golubitsky2002}. The full dynamics, which are 
described by a set of PDEs, can be reduced in the neighbourhood of the 
bifurcation point to simple ODEs of the form
\begin{equation}
\dd{\boldsymbol{y}}{t} = \boldsymbol{f}\left(\boldsymbol{y}; \mu\right)  
\end{equation}
using methods such as centre-manifold reduction or Lyapunov-Schmidt reduction. 
Such equations are termed \textit{amplitude equations}. $\boldsymbol{y}$ is the 
vector of mode amplitudes (which would be referred to as an \textit{order 
parameter} in crystallography). The vector $\boldsymbol{y}$ is of the same 
dimension 
as the irrep.  $\mu$ is the bifurcation parameter, which for 
convection problems can be related to the Rayleigh number. Bifurcation occurs 
when $\mu$ passes through zero.

The function $\boldsymbol{f}\left(\boldsymbol{y}; \mu\right)$ is 
\textit{equivariant} 
under the 
action of the matrices of the given irrep, that is
\begin{equation}
 \boldsymbol{f}\left( \mathsfbi{g} \boldsymbol{y}; \mu\right) = \mathsfbi{g}  
\boldsymbol{f}\left(\boldsymbol{y}; 
\mu \right) 
\end{equation}
for all matrices $\mathsfbi{g}$ in the given irrep. Equivariance places 
strong 
constraints on the form of the amplitude equations, and in turn on the nature 
of 
the bifurcation. Equivariance under a non-trivial irrep implies that 
$\boldsymbol{f}\left(\boldsymbol{0}; \mu\right) = 
\boldsymbol{0}$ and hence $\boldsymbol{y} = \boldsymbol{0}$ is always an 
equilibrium solution (although 
not 
necessarily a stable one). Steady-state bifurcations without symmetry 
constraints are generically saddle-node bifurcations; it is the 
constraints from symmetry that lead to pitchfork or transcritical 
bifurcations instead. 

The simplest example of the consequences of equivariance are in a 1-dimensional 
system, invariant under $C_2 = \lbrace 1, -1 \rbrace$. Equivariance under $C_2$ 
implies the function $f$ is odd ($f(-y; \mu) = - f(y; \mu)$) which in turn 
implies that in a Taylor expansion of $f(y; \mu)$ about $y=0$ no even-order 
terms in $y$ will appear. It follows from the symmetry alone that the 
associated bifurcation must be a pitchfork.

A common method of analysing amplitude equations is to consider their Taylor 
expansion in powers of $\boldsymbol{y}$, and to truncate at some particular 
order. 
Much generic behaviour about the bifurcation can be described by these 
truncated 
forms. Moreover, symmetry places constraints on the number of independent 
parameters needed to describe the truncated form: for the $C_2$ example there 
are no quadratic or other even order terms present. The dimension of the space 
of equivariants of given degree can be obtained purely using the characters of 
the representation (see Appendix \ref{sec:equi} and \citet{Antoneli2008}). This 
can be used to show for example that the faithful irrep of $D_3$ has a 
quadratic 
equivariant, unlike $C_2$. The faithful irrep of $D_3$ is generically 
associated with a transcritical 
bifurcation, although for a particular problem there is always the possibility 
that the coefficient associated with the quadratic equivariant is zero due to 
some particular feature of the governing equations (e.g. self-adjointness, 
\citet{Golubitsky1984}) that 
would then cause the bifurcation to be a pitchfork.

The final column of \autoref{tab:maxsub_p4nmm} classifies the type of generic 
steady-state bifurcation associated with each of the maximal subgroups of 
$p4/nmm$. Only the index-9 k-transition is associated with a transcritical 
bifurcation; all others are pitchforks. The bifurcations can generically be 
classified depending on whether the dynamics when restricted to the fixed-point 
subspace has a quadratic term: pitchfork if not, transcritical if so. The 
classification of bifurcations is discussed further in supplement 3, which 
provides character tables of several small finite groups and the dimensions of 
their spaces of equivariants.

\section{Convection}\label{sec:conv}

\begin{figure}
 \includegraphics[width=\columnwidth]{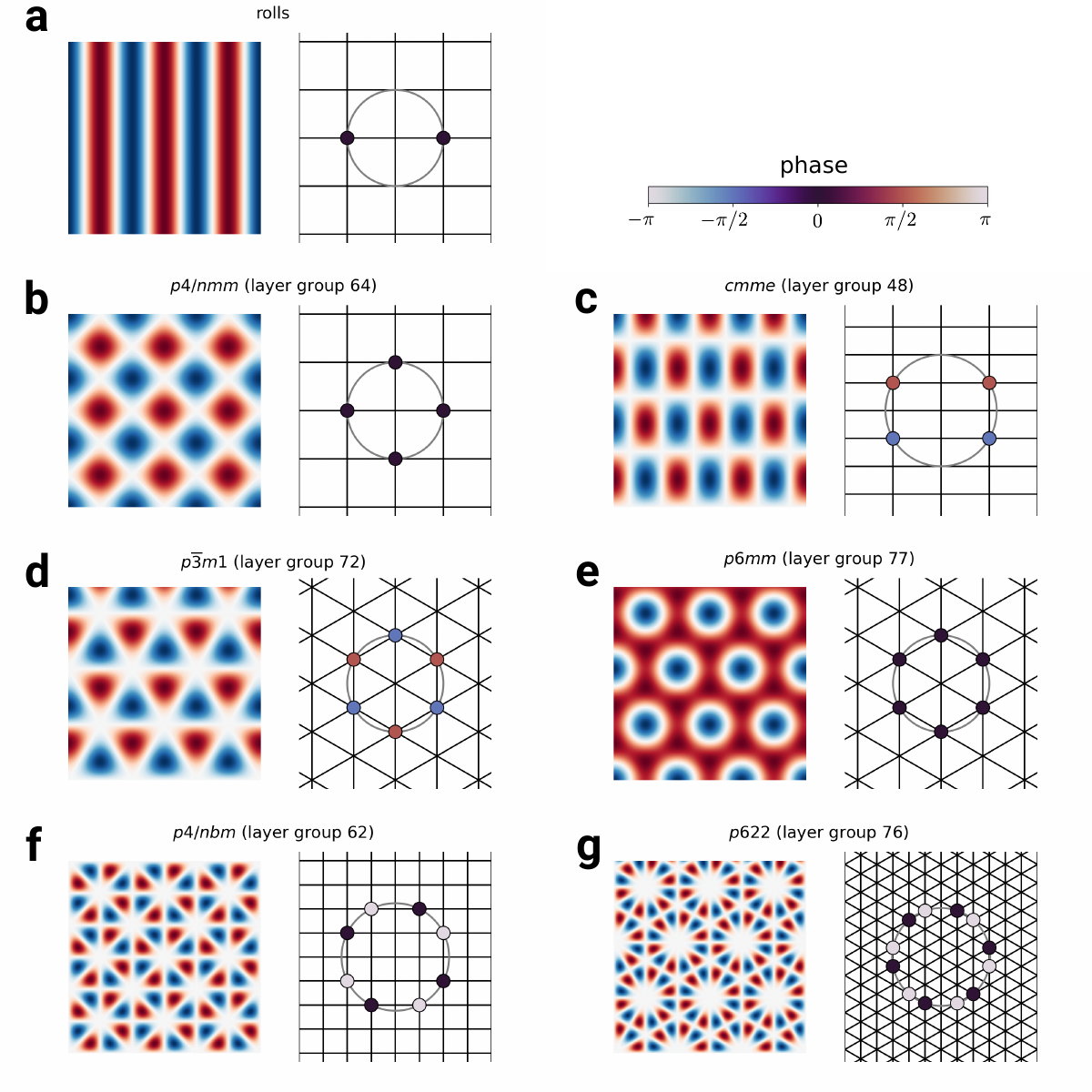}
 \caption{Examples of crystallographic classification for convective flows 
consisting of a single horizontal wavenumber. Shown are a) rolls, b) squares 
(checkerboard), c) rectangles (patchwork quilt), 
d) triangles, e) down-hexagons, f) anti-squares, g) 
anti-hexagons. Each pattern, with the exception of rolls, is 
labelled by its Hermann-Mauguin layer group symbol. The pattern of rolls does 
not correspond to a layer group, as it has one axis with a continuous 
translation symmetry (its symmetry may be referred to as $\mathcal{p}_a 
\mathcal{\nu}_b ma2$, \citet{Kopsky2006}). 
The left plot of each panel shows the mid-plane 
temperature field, the right plot shows its Fourier transform (reciprocal space 
plot). In reciprocal space the size of the dots show the amplitude, the colour 
of the dots show the phase (colourbar in top right). Grid lines indicate the 
reciprocal lattice, although note that some mode patterns are consistent 
with more than one type of lattice (e.g. both hexagonal and rectangular). The 
lattice shown is that used in ITE for the given layer group. With a single 
horizontal wavenumber all modes must lie on a circle in reciprocal space 
(grey line). All of the above patterns represent a single parameter family: 
once the origin and orientation is specified the only remaining parameter is 
the amplitude.}
\label{fig:examples1}
\end{figure}

We will now apply the theory discussed in the previous sections to transitions 
in fluid layers, and in particular to thermal convection. Consider a layer of 
fluid, heated from below and cooled from above. As the Rayleigh number is 
increased past some critical value the system begins to convect. Depending on 
choices of boundary conditions and rheology different planforms of the flow are 
possible: common planforms seen at onset are rolls, hexagons, and squares. Each 
of these convective planforms can be classified using crystallographic 
notation, 
e.g. squares have layer group symmetry $p4/nmm$ (layer group 64), as 
illustrated 
in \autoref{fig:examples1}b.

The physical state of the fluid can be described by its temperature field. 
\autoref{fig:examples1} shows examples of possible temperature fields that can 
occur at the initial onset of convection. Each panel shows the mid-plane 
temperature field with red/blue colouring for hot/cold, along with the 
corresponding reciprocal space (Fourier domain) pattern, where each dot is 
coloured according to phase, and the size of the dot indicates its amplitude. 
At 
the onset of convection there is typically a single critical horizontal 
wavenumber $k_c$, and the horizontal variation is described by a planform 
function $f(x,y)$ satisfying $\nabla_h^2 f = - k_c^2 f$ \citep{Ribe2018}. When 
constrained to a periodic lattice, the planform function is a superposition of 
modes of the form $\e^{i (k_x x + k_y y)}$ where $\boldsymbol{k} = (k_x, k_y)$ 
is 
the horizontal wavenumber vector and $|\boldsymbol{k}|=k_c$. Thus in the 
reciprocal 
space plots of \autoref{fig:examples1} all the dots lie on a circle of radius 
$k_c$.

Panels f and g in \autoref{fig:examples1} show examples of superlattice 
patterns \citep{Hoyle2006, Dionne1997, Dawes2003}. These are patterns where the 
critical 
wavenumber $k_c$ is larger in magnitude than the basis vectors describing the 
reciprocal lattice. For example, in \autoref{fig:examples1}f the basis vectors 
of the reciprocal lattice are $\boldsymbol{k}_1 = (0, 1)$ and $\boldsymbol{k}_2 
= (1, 0)$ and the critical circle has $k_c = \sqrt{5} > 1$. Superlattice 
patterns show periodicity on one scale (in \autoref{fig:examples1}f periodicity 
in $x$ and $y$ with a period of $2 \pi$), but features in the pattern occur on 
a smaller scale (in \autoref{fig:examples1}f with a wavelength of $2 \pi / 
\sqrt{5}$.)

\begin{figure}
 \includegraphics[width=\columnwidth]{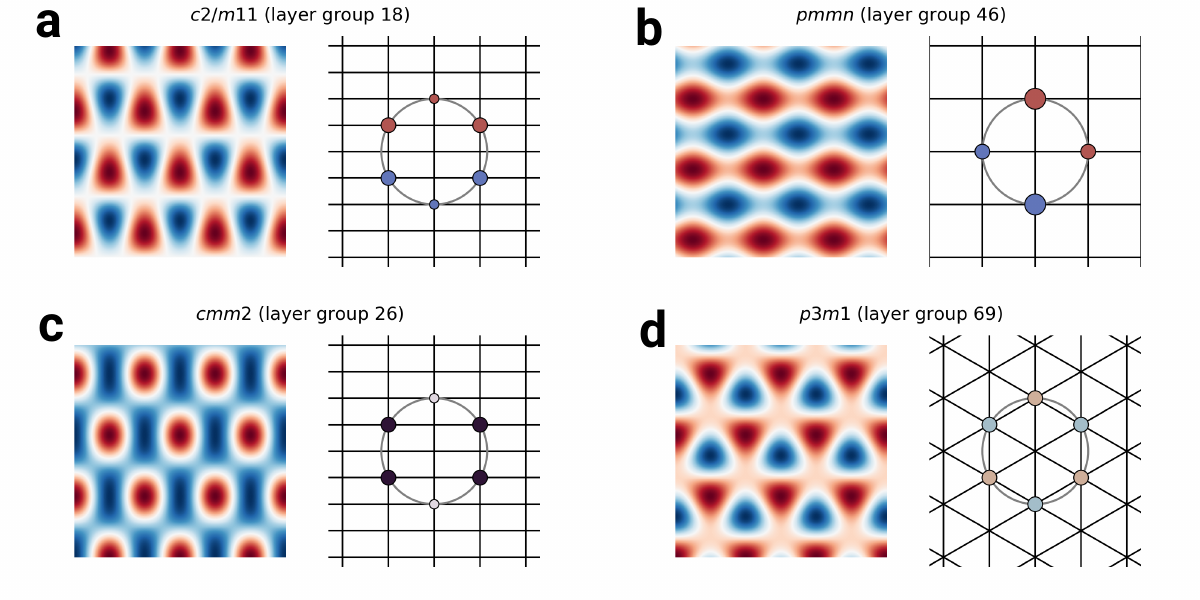}
 \caption{Further examples of crystallographic classification for convective 
flows consisting of a single horizontal wavenumber. These examples form
two parameter families, and each pattern may be considered as a superposition 
of two of the single parameter patterns shown in \autoref{fig:examples1}. 
Shown are a) trapezoids (a combination of squares (64) and triangles (72)) b) 
bimodal (a combination of squares (64) and 
rolls or two orthogonal sets of rolls)  c) up-rectangles (a combination of 
rectangles (48) and hexagons(77)), d) down-triangles (a combination of 
triangles (72) and hexagons (77))}
\label{fig:examples2}
\end{figure}

Each of the 
patterns illustrated in \autoref{fig:examples1} is a single parameter family: 
once the origin and orientation of the pattern are specified, the only 
remaining 
parameter that describes the flow is the amplitude. \autoref{fig:examples2} 
illustrates two-parameter examples that still have a single horizontal 
wavenumber (e.g. bimodal flow). The initial onset of convection and the 
selection of convective planform has been very well studied (see e.g. the 
extensive studies by \citet{Buzano1983, Golubitsky1984, Knobloch1990}); we 
simply note here that each of the convective planforms which are typically 
described in the convective literature by a name (like hexagons, bimodal flow, 
patch-work quilt etc.) can be given a Hermann-Mauguin symbol that unambiguously 
specifies its symmetry. 

\subsection{Numerical simulations}

As the Rayleigh number is increased the initial convective planforms of 
hexagons, rolls, squares etc. undergo a series of further symmetry-breaking 
transitions. Typically such transitions are investigated using numerical 
simulations. Ideas from group theory can both illuminate the results of the 
numerical simulations and be used to make the computations more efficient.

As a concrete example, consider a 3-dimensional numerical simulation of 
fixed-flux convection in a constant-viscosity fluid layer at infinite Prantl 
number with the Boussinesq approximation (the governing equations can be found 
in appendix \ref{sec:rb_gov_eq}). At the onset 
of convection the 
expected planform is squares \citep{Proctor1981}, so it is natural to consider 
a computational domain that is a box with periodic boundary conditions 
in the 
horizontal. The temperature field within the box is described in terms of 
coefficients with respect to some finite set of basis vectors. The particular 
calculations here use spectral basis elements of the form
\begin{equation}
 \theta(x,y, z) = \sum_{k = -K}^K \sum_{l = -L}^L \sum_{m=0}^{M} c_{klm} \e^{\i 
(k 
x + l y)} T_m(z)
\end{equation}
i.e. a basis of Fourier modes in the horizontal, and Chebyshev polynomials in 
the vertical \citep{Burns2020}. However, the same group theory ideas can be 
exploited whatever 
choice of basis is made. Since $\theta$ (the temperature perturbation) is a 
real 
variable, $c^*_{klm} = c_{\overline{k}\overline{l}m}$.

Suppose there is a $N$-dimensional set of coefficients describing the given 
state. Each symmetry can be represented by a $N$-by-$N$ matrix which describes 
the action of that symmetry on the basis coefficients (here the set of 
$c_{klm}$). In general this $N$-by-$N$ representation is reducible, and it is 
possible to change basis such that in the new basis the components transform 
according to the irreducible representations of the given group.

The change of basis is achieved using projection operators. To project onto the 
components which transform according to the $J$-th irrep of a group $G$ we 
apply the operator $\mathsfbi{P}^J$ defined by
\begin{equation}
 \mathsfbi{P}^J = \frac{\operatorname{dim} J}{| G| } \sum_{g \in G} (\chi^J 
(g))^* 
\mathsfbi{g} 
\end{equation}
where $\operatorname{dim} J$ is the dimension of the irrep, $| G|$ is the order 
of the group, $\chi^J(g)$ is the 
character of the element $g$, and $\mathsfbi{g}$ is the matrix representing the 
action of the element $g$ in the given representation. Moreover, it should be 
noted that through a change of basis the representations can be made unitary 
(orthogonal in the case of real representations) using Weyl's unitary trick. In 
turn, an orthogonal projection matrix can be used to give a orthogonal set of 
basis 
vectors corresponding to a particular irrep using a QR decomposition.  

\begin{figure}
 \includegraphics[width = \columnwidth]{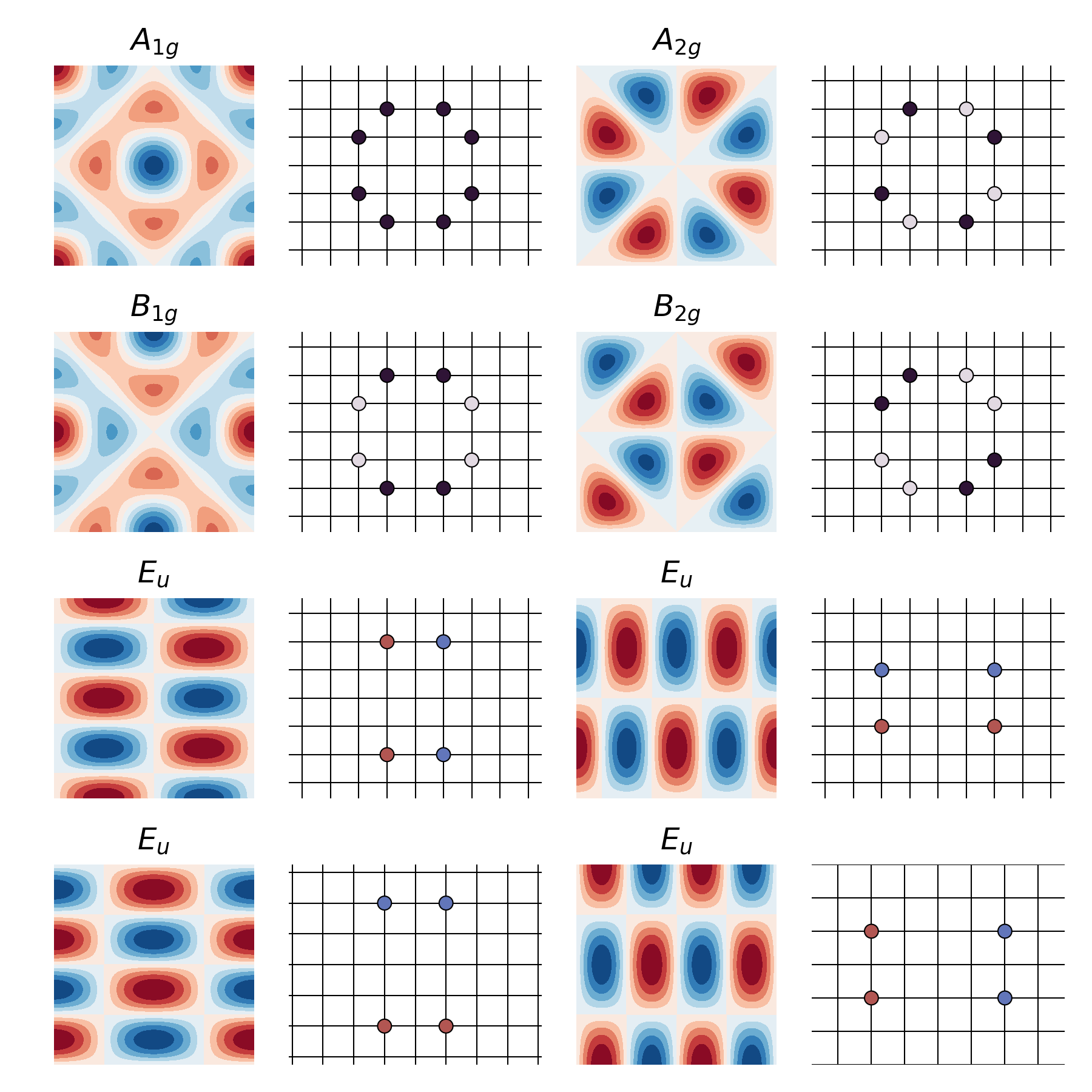}
 \caption{An example of an isotypic decomposition for the wavevector star 
generated by $\boldsymbol{k} = (2, 1)$ with no $z$-dependence. Shown is the 
decomposition into the irreps 
of $p4/nmm$ given in \autoref{tab:charp4nmm}. The star decomposes as $A_{1g} 
\oplus A_{2g} \oplus B_{1g} \oplus B_{2g} \oplus 2 E_u$.}
\label{fig:isotypic}
\end{figure}

An illustration of an isotypic decomposition into basis vectors which transform 
according to the irreps is given in \autoref{fig:isotypic}. This example 
considers the $c_{210}$ coefficient, and the coefficients to which it can be 
related using the layer symmetry $p4/nmm$. The periodicity of the computational 
domain is assumed to align with the principal lattice of translations of 
$p4/nmm$. As such the basis vectors $c_{klm}$ are invariant under the group $T$ 
of lattice translations, so one only needs to consider the factor group $G/T$ 
whose character table is given in \autoref{tab:charp4nmm}. There are 8 
coefficients that are related by symmetry to $c_{210}$, and the 8 dimensional 
space can be decomposed into the irreps given in \autoref{tab:charp4nmm} as the 
direct sum $A_{1g} 
\oplus A_{2g} \oplus B_{1g} \oplus B_{2g} \oplus 2 E_u$ (see \citet{Dionne1997} 
for an application of this particular decomposition). 

\begin{figure}
 \includegraphics[width=1.1\columnwidth]{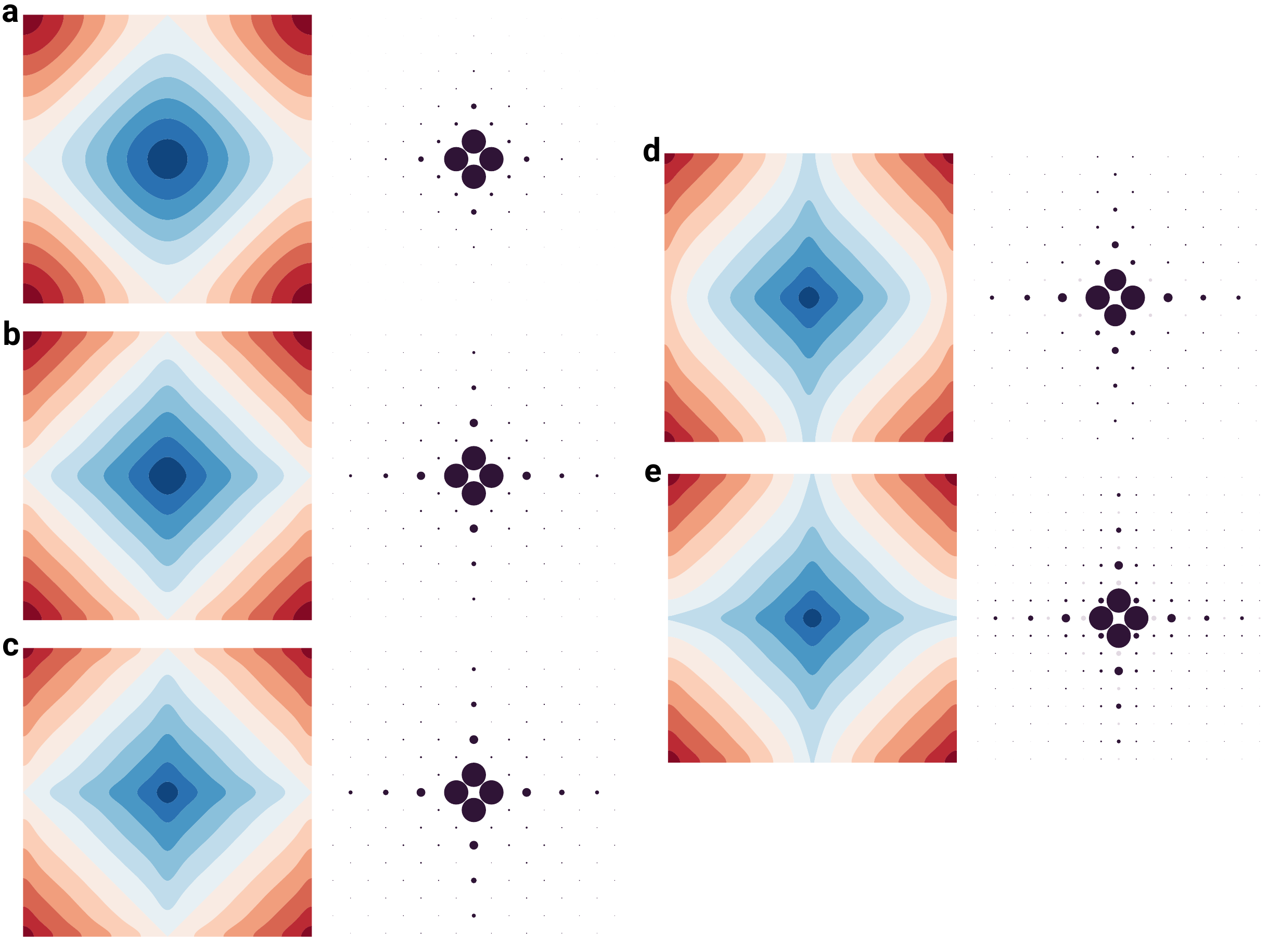}
 \caption{Examples of symmetry-breaking bifurcations in fixed-flux convection 
in a fluid layer. Shown are images of the mid-plane temperature field, both in 
real space (contour plots) and in reciprocal space (dot patterns). At the onset 
of convection a square planform is seen, with $p4/nmm$ symmetry (layer group 
64). The left panels show the evolution of the $p4/nmm$ solution as the 
Rayleigh number is increased from a) $\Ra = 200$, b) $\Ra = 700$, 
and c) $\Ra = 1500$. The $p4/nmm$ solution at $\Ra = 
1500$ is unstable to perturbations that break the symmetry. Panels d) and e) 
show new solutions at $\Ra = 
1500$ that emerge from pitchfork bifurcations from the $p4/nmm$ 
solution. d) has symmetry $pmmn$ (layer group 46) and e) has symmetry $p4mm$ 
(layer group 55).}
\label{fig:convection_break}
\end{figure}

\begin{figure}
\centering
 \includegraphics[width=0.63\columnwidth]{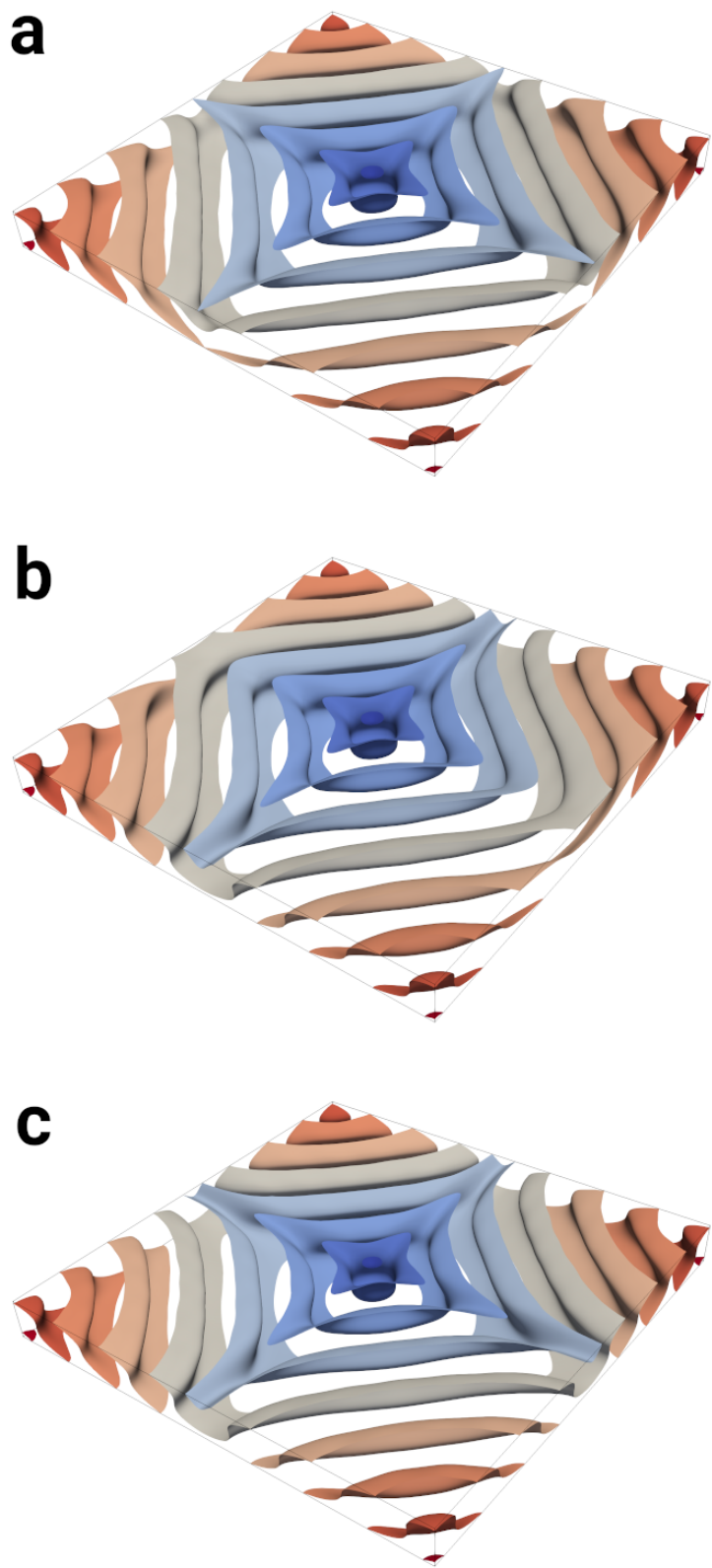}
 \caption{3-dimensional rendering of the convective flows shown in 
horizontal cross-section in \autoref{fig:convection_break}c, d, e. Shown are 
equally-spaced contours of the temperature field. All flows have $\Ra = 
1500$. a) has symmetry $p4/nmm$, b) has symmetry $pmmn$, c) has symmetry 
$p4mm$. 
The loss of the 4-fold vertical rotation symmetry $4_z$ about the centre of 
the box in going from a to b 
can 
be clearly seen. The loss of symmetry between a and c is more subtle: the 
upwellings are now not related by symmetry to the downwellings. c has lost the 
horizontal glide reflection $n$, the 2-fold rotations about horizontal axes, 
and 
the 2-fold screw rotations about horizontal axes (see \autoref{fig:p4nmm_sym}).}
\label{fig:3drenders}
\end{figure}

\begin{figure}
\centering
 \includegraphics[width=\columnwidth]{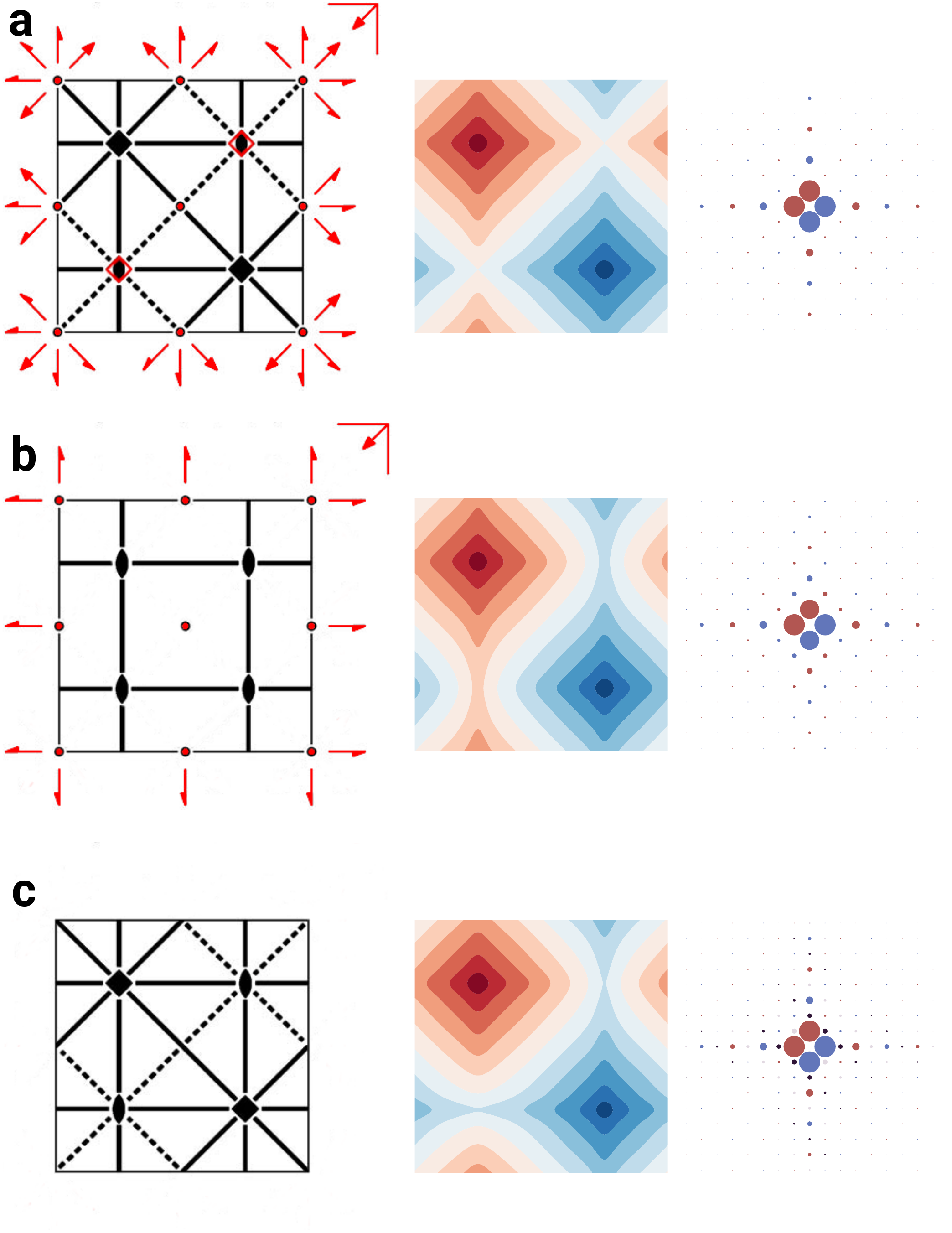}
 \caption{Plots identical to \autoref{fig:convection_break}c, d, e but with 
the origin of the coordinate system shifted by $(\tfrac{1}{4}, \tfrac{1}{4}, 
0)$ (the coordinate system given as origin choice 2 for $p4/nmm$ in ITE). 
The corresponding symmetry diagrams (with origin shifted from 
\autoref{fig:p4nmm_sym}) are shown on the left. 
All flows have $\Ra = 
1500$. a) has symmetry $p4/nmm$, b) has symmetry $pmmn$, c) has symmetry 
$p4mm$. Some of the symmetry losses are clearer to see with this 
choice of origin as the rotation axes are moved away from the edges of the box.
The loss of the 4-fold inversion axes ($\overline{4}_z$) in going 
from a to b or c can be clearly seen. The temperature perturbation is 
necessarily zero on the mid-plane at a 4-fold inversion axis.}
\label{fig:shift}
\end{figure}

The isotypic decomposition can be used to simplify the numerical study of 
symmetry-breaking bifurcations. An example of this is shown in Figures
\ref{fig:convection_break}, \ref{fig:3drenders} and \ref{fig:shift} which 
illustrates breaking of a $p4/nmm$ pattern 
of squares into two different planforms with less symmetry, one of $pmmn$ and 
one of $p4mm$. The computational domain is a box, with aspect ratio such that 
the distance between rising and sinking regions is 8 times the layer depth. The 
heat flux is fixed on the top and bottom boundaries and both boundaries are 
free-slip. Fixed-flux convection with free-slip boundaries in an infinite layer 
formally has $k=0$ as the most unstable wavenumber, with critical Rayleigh 
$\Ra_c = 120$ \citep{Chapman1980a,Rieutord2015}. For the finite 
horizontal 
scale of the numerical problem the critical Rayleigh number is slightly higher, 
$\Ra_c = 126$. \autoref{fig:convection_break}a shows the planform near 
onset, at $\Ra=200$, which is dominated by the four modes on the critical 
circle $|\boldsymbol{k}| = k_c$ although there are also small contributions 
from 
higher modes.

As the Rayleigh number is increased there is more power in higher modes 
(\autoref{fig:convection_break}b, c) and sharper features are seen. However, 
the 
solution shown in \autoref{fig:convection_break}c (and also 
\autoref{fig:3drenders}a and \autoref{fig:shift}a) at $\Ra=1500$ is 
actually unstable to perturbations that break the symmetry. The unstable 
solution can be computed by imposing the $p4/nmm$ symmetry on the numerical 
scheme, either by restricting the set of basis vectors used to those associated 
with the trivial irrep $A_{1g}$, or by projecting the solutions onto that irrep 
at each iteration using the projection operator. Restricting the set of basis 
vectors used is more advantageous in terms of computational efficiency, as one 
then solves a problem with a much smaller set of unknowns.

Figures \ref{fig:convection_break}, \ref{fig:3drenders} and \ref{fig:shift} 
illustrate the same symmetry-breaking transitions, but do so using different 
projections of the numerical solutions. Figures 
\ref{fig:convection_break}c, \ref{fig:3drenders}a and \ref{fig:shift}a show the 
unstable $p4/nmm$ solution; Figures 
\ref{fig:convection_break}d, \ref{fig:3drenders}b and \ref{fig:shift}b show the 
symmetry break to a $pmmn$ solution; Figures 
\ref{fig:convection_break}e, \ref{fig:3drenders}c and \ref{fig:shift}c show the 
symmetry break to a $p4mm$ solution. Both \autoref{fig:convection_break} and 
\autoref{fig:shift} show the mid-plane plane temperature field and its 
corresponding reciprocal space plot: the only difference is the choice of 
origin. \autoref{fig:convection_break} uses what is given in ITE as origin 
choice 1 for $p4/nmm$, the same as used in the symmetry diagrams of 
\autoref{fig:p4nmm_sym}. \autoref{fig:shift} uses an origin that is shifted by 
$(1/4, 1/4, 0)$ and referred to as origin choice 2 in ITE. 
\autoref{fig:3drenders} give a 3d-rendering of the isotherms in origin 
choice 1 co-ordinates.

The loss of symmetry is perhaps clearest to see in the origin choice 2 
mid-plane images of \autoref{fig:shift}. The loss of the four-fold rotation 
axes in going from \autoref{fig:shift}a ($p4/nmm$) to \autoref{fig:shift}b 
($pmmn$) is particularly apparent. The loss of symmetry in going from 
\autoref{fig:shift}a ($p4/nmm$) to \autoref{fig:shift}c ($p4mm$) is more subtle 
as it involves the loss of the glide reflection about the horizontal mid-plane. 
The upwellings in \autoref{fig:shift}c are no longer related by symmetry to the 
downwellings as they are in \autoref{fig:shift}a. This change can be seen in 
the differences between the shape of the blue contours (cold downwellings) and 
the red contours (hot upwellings) in \autoref{fig:shift}c: there are narrow 
connections between the blue downwelling regions, but no connections 
between the red upwelling regions. These connections can also be seen in the 3d 
rendering of \autoref{fig:3drenders}c coming through the middle of each side of 
the box as light blue contours.

The stability of the $p4/nmm$ solution as a function of Rayleigh number can be 
assessed using standard linear stability analysis, but the calculations can be 
made more efficient by exploiting the symmetry. A good general introduction to 
the numerical methods for performing linear stability and bifurcation analysis 
can be found in \citet{Tuckerman2000}. The stability analysis relies on the 
calculation of the eigenvalues of an appropriate Jacobian matrix. When 
an eigenvalue has a real part which goes from being negative to being 
positive there is instability and an associated bifurcation to a new flow 
pattern. The isotypic decomposition aids the linear stability analysis by 
allowing one to block-diagonalise the Jacobian according to the irreps. This 
has 
several advantages: i) there are smaller linear systems to deal with in the 
individual blocks; ii) the eigenvalues in the individual blocks may be more 
widely separated that those of the full problem, speeding up convergence of 
numerical eigenvalue techniques; iii) one can directly identify the symmetries 
that are broken and the corresponding active irrep. 

\begin{figure}
\centering
 \includegraphics[width=0.7\columnwidth]{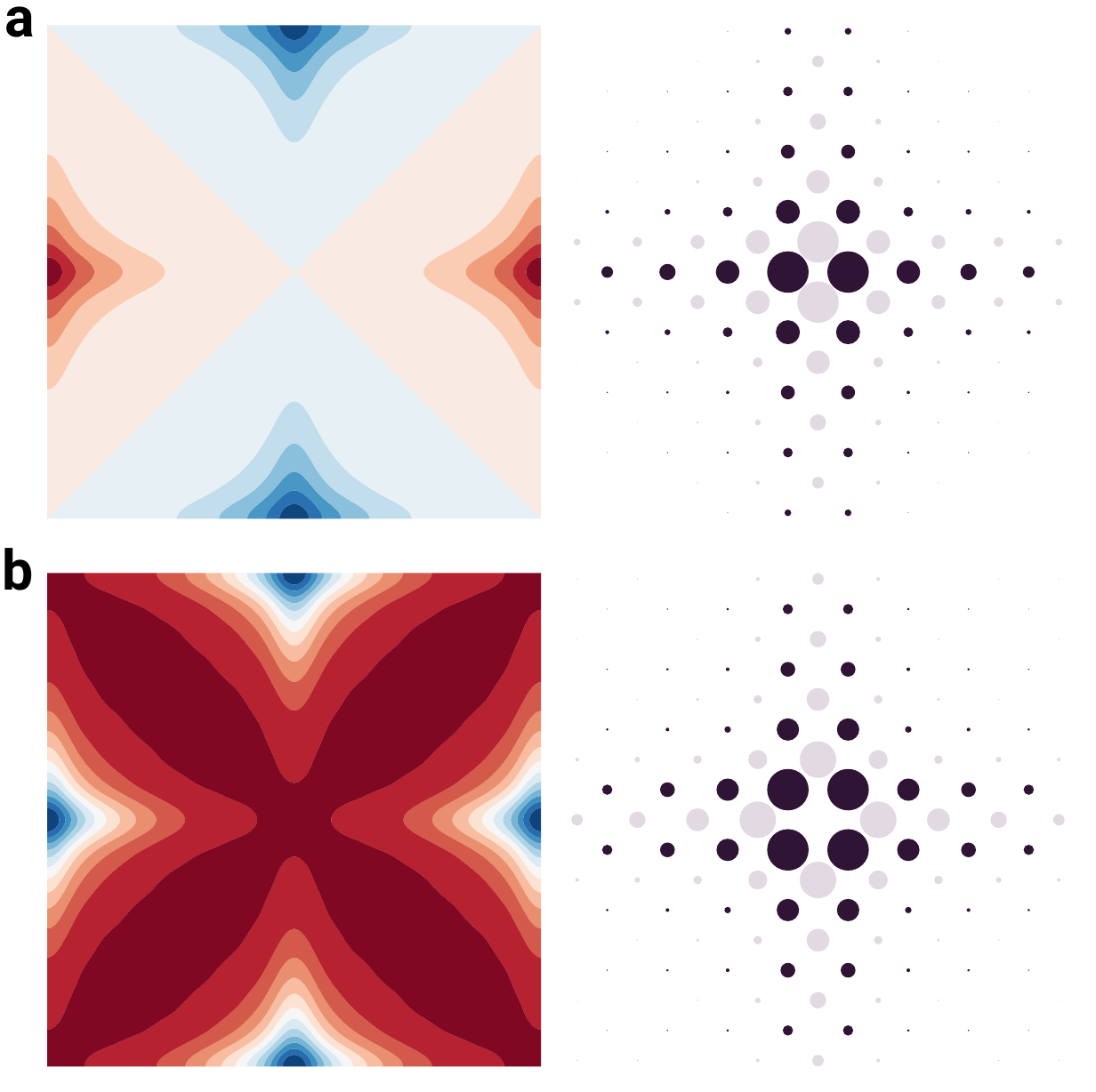}
 \caption{Examples of eigenmodes in a linear stability analysis of the $p4/nmm$ 
solution depicted in \autoref{fig:convection_break}b with $\Ra=700$ (using 
origin choice 1). a) shows 
the mid-plane 
temperature field for the eigenmode with eigenvalue with largest real part 
which transforms according to the irrep $B_{1g}$ in \autoref{tab:charp4nmm}, 
associated with the bifurcation to the $pmmn$ solution in 
\autoref{fig:convection_break}d. b) shows the corresponding eigenmode which 
transforms according to irrep $A_{2u}$ in \autoref{tab:charp4nmm}, associated 
with bifurcation to $p4mm$ in \autoref{fig:convection_break}e}
\label{fig:eigenmodes}
\end{figure}

\autoref{fig:eigenmodes} illustrates the linear stability analysis of the 
$p4/nmm$ solution in \autoref{fig:convection_break}b at $\Ra=700$. Shown 
are the eigenmodes with largest real part corresponding to the irreps $B_{1g}$ 
and $A_{2u}$. At $\Ra=700$, eigenvalues of both modes are real and 
negative. However, for slightly larger $\Ra$, at $\Ra=756$ for 
$B_{1g}$ and $\Ra=815$ for $A_{2u}$, the eigenvalues become positive, 
leading to bifurcations and the solutions with broken symmetry seen in 
\autoref{fig:convection_break}d, e (and also \autoref{fig:3drenders}b, c and 
\autoref{fig:shift}b, c). Given the irreps involved, these 
bifurcations are necessarily pitchfork bifurcations.

Symmetry can also be exploited to calculate the new equilibrium states after a 
bifurcation. The solution without the symmetry-break can be used as an 
initial condition, with a 
small perturbation added in the form of the symmetry-breaking eigenmode of the 
linear stability calculation. One can use the projection operators to constrain 
the solution to have the appropriate symmetry (e.g. for 
\autoref{fig:convection_break}d, imposing the group $pmmn$ or restricting to 
only those 
basis vectors corresponding to the irreps $A_{1g}$ and $B_{1g}$ of $p4/nmm$ in 
\autoref{tab:charp4nmm}). Once the eigenmodes associated with the bifurcations 
have been calculated it is possible to systematically perform a centre manifold 
reduction to determine the amplitude equations \citep{Carini2015}, although we 
have not done this here.

\subsection{Further examples}

There are many examples of flow transitions in fluid layers in the literature, 
but almost none use the crystallographic notation that has been adopted here. 
One study that does is \citet{McKenzie1988}, which describes a wide variety of
transitions in convection, particularly those in the experimental studies 
of a temperature-dependent viscosity fluid by 
\citet{White1988a}. A convective system with a temperature-dependent viscosity 
is not invariant under reflection in a horizontal mirror plane. As such the 
layer groups involved are those without such mirror planes, which are 
equivalent to those of the 17 plane or wallpaper groups. Many of the examples 
discussed by \citet{McKenzie1988} are pitchfork bifurcations, although it 
should be noted that those bifurcations he discusses with factor group $D_3$ 
are generically 
transcritical, and not pitchforks.

\section{Conclusions}

Our aim in this work has been to demonstrate the utility of the extensive 
databases in crystallography for understanding transitions in fluid layers. 
For simplicity, we have focussed on steady-states, doubly-periodic in space 
that are described by the crystallographic layer groups. We have not discussed 
transitions which break the time-translation symmetry (Hopf 
bifurcations) which involve spatio-temporal group elements that combine space 
group elements with time translations. The bifurcation theory for such cases is 
well understood, but it would be helpful to have a 
standardised crystallographic notation for describing such transitions. 
Translations in time can be associated with a fourth dimension. Hopf 
bifurcations lead to solutions that are time-periodic, and thus the natural 
groups to consider will be the subperiodic groups in four-dimensions with a 
three-dimensional lattice of translations (the two horizontal space dimensions 
and one time dimension). We have 
also made no attempt here to describe the representation theory for the initial 
onset of convection: formally this would involve a study of the representations 
of $E(2) \times C_2$ which is a non-compact Lie group. The difficulty of 
dealing with such a group is usually side-stepped in the literature by 
considering instead a problem on a compact domain (the two-torus). The analysis 
of pattern-forming problems is considerably more straightforward 
when they are forced to be periodic with respect to a lattice as they have been 
in this manuscript. The spectra of modes is discrete and techniques like 
centre manifold reduction can be used. Patterns which are not exactly periodic 
are more challenging to analyse, but there are techniques available (see 
for example Chapters 7 to 11 of the textbook by \citet{Hoyle2006}). Such 
methods are required for understanding phenomena such as defects, 
dislocations, the zig-zag and cross-roll instabilities, spirals, and 
quasipatterns.   

There is a wealth of useful information that lies within the crystallographic 
databases, and we encourage fluid dynamicists to exploit it.

\bibliography{groups}
\bibliographystyle{jfm}

\backsection[Supplementary data]{\label{SupMat}Three supplements with 
additional group theory tables are available at 
\\https://doi.org/10.1017/jfm.xx}

\backsection[Acknowledgements]{We thank Jonathan Dawes, Alastair 
Rucklidge and three anonymous reviewers for their helpful comments on this 
work.}

\backsection[Funding]{This research received no specific grant from any funding 
agency, commercial or not-for-profit sectors.}

\backsection[Declaration of interests]{The authors report no conflict of 
interest.}




\appendix

\section{The symmetries of Rayleigh-B\'{e}nard convection}\label{sec:rb_sym}

Consider Rayleigh-B\'{e}nard convection in a fluid layer, with $x$ and $y$ as 
horizontal 
coordinates and $z$ as a vertical coordinate. The system has a natural 
Euclidean symmetry in the horizontal plane, represented by the group $E(2)$. 
However, depending on boundary conditions and rheological choices there may be 
additional symmetries in the problem.

\subsection{Governing equations}\label{sec:rb_gov_eq}

For Boussinesq, infinite Prantl number, thermal convection the governing 
equations are
\begin{gather}
\bnabla \bcdot \v = 0, \label{eq:g1} \\
\bnabla \bcdot \sig = - \rho_0 g \alpha T \hat{\boldsymbol{z}},  \label{eq:g2} 
\\
\pdd{T}{t} + \v \bcdot \bnabla T = \kappa \nabla^2 T,  \label{eq:g3}
\end{gather}
where $\v$ is the fluid velocity, $\sig$ is the stress tensor, $\rho_0$ is the 
reference density, $g$ is the acceleration due to gravity, $\alpha$ is the 
thermal expansivity, $T$ is the temperature, and $\kappa$ is the thermal 
diffusivity. The Newtonian constitutive law relating stress to strain rate 
is
\begin{equation}
 \sig = - p \mathsfi{I} + \eta \left(\bnabla \v + \bnabla \v^\mathrm{T} \right),
\end{equation}
where $p$ is the pressure. Let $\theta$ represent the temperature perturbation 
from a conductive steady 
state, where the steady-state temperature gradient is $\Delta T / a$, and $a$ 
is the layer thickness. The governing equations \eqref{eq:g1}, \eqref{eq:g2}
and \eqref{eq:g3} can be rewritten as
\begin{gather}
\bnabla \bcdot \v = 0, \\
\bnabla \bcdot \tilde{\sig} = - \rho_0 g \alpha \theta \hat{\boldsymbol{z}}, \\
\pdd{\theta}{t} + \v \bcdot \bnabla \theta - \frac{\Delta T}{a} \v \bcdot 
\hat{\boldsymbol{z}} = \kappa \nabla^2 \theta,
\end{gather}
where $\tilde{\sig}$ is a modified stress tensor which represents the 
difference from the conductive state. The equations can be made dimensionless 
be scaling all lengths by the layer thickness $a$, and all times by the 
diffusion time $a^2/\kappa$. The behaviour is controlled by the 
dimensionless Rayleigh number $\Ra = \rho_0 g \alpha \Delta T 
a^3 / (\eta_0 \kappa )$. The temperature can be scaled by $\Delta T / 
\Ra$, the velocity by $\kappa/a$, and the pressure by 
$\rho_0 g \alpha \theta_0 d$ to yield
\begin{gather}
\bnabla \bcdot \v = 0, \\
- \bnabla \bcdot \tilde{\sig} = \theta \hat{\boldsymbol{z}}, \\
\frac{1}{\Ra} \left( \pdd{\theta}{t} + \v \bcdot \bnabla \theta - 
\nabla^2 \theta \right)  = \v \bcdot \hat{\boldsymbol{z}}.
\end{gather}

\subsection{Mid-plane reflection}

If boundary conditions top and bottom are identical, then providing the 
viscosity is constant (or depth-dependent with mid-plane symmetry), the 
equations are invariant under the mid-plane symmetry ($z=0$ 
is the 
mid-plane),
\begin{equation}
 m_z: (x, y, z) \rightarrow (x, y, -z), \label{eq:euc_4_again}
\end{equation}
provided the variables in the equations transform as
\begin{align}
 m_z: \quad& u \rightarrow u, \quad v \rightarrow v, \quad p \rightarrow p, 
\nonumber \\
 &\theta \rightarrow - \theta, \quad w \rightarrow - w \label{eq:varchange}
\end{align}
where the velocity vector is $\boldsymbol{v} = (u, v, w)$ and $p$ is the 
pressure 
perturbation. \eqref{eq:varchange} represents the symmetry between hot 
upwellings and cold downwellings.


\subsection{Poloidal-toroidal decomposition}

For 3D flows represented by poloidal and toroidal potentials $\mathcal{S}$ and 
$\mathcal{T}$ (and neglecting any mean-field flow)
\begin{gather}
 \v = \bnabla \times (\zhat \times \bnabla \mathcal{S}) + \zhat \times  \bnabla 
\mathcal{T}, \nonumber \\
 u = - \frac{\partial^2 \mathcal{S}}{\partial x \partial z} - \frac{\partial 
\mathcal{T}}{\partial y}, \quad  v = - \frac{\partial^2 \mathcal{S}}{\partial y 
\partial z} + \frac{\partial \mathcal{T}}{\partial x}, \quad w = - \nabla^2_h 
\mathcal{S}
\end{gather}
so the potentials must transform under mid-plane reflection as
\begin{equation}
  m_z: \quad \mathcal{S} \rightarrow - \mathcal{S}, \quad \mathcal{T} 
\rightarrow \mathcal{T}.
\end{equation}
This is different from their transformation under vertical mirrors, which is  
\begin{equation}
  m_x: \quad \mathcal{S} \rightarrow \mathcal{S}, \quad \mathcal{T} 
\rightarrow - \mathcal{T}.
\end{equation}
as $m_x: u \rightarrow - u$. For the constant-viscosity example in 
\autoref{fig:convection_break} the flow is purely poloidal 
($\mathcal{T}=0$). The poloidal potential $\mathcal{S}$ transforms in the same 
way as the temperature perturbation $\theta$ under the symmetry operations. 

While this work uses the Cartesian co-ordinates of the plane layer, 
it should be noted that poloidal-torodial decompositions are commonly used in 
problems of spherical geometry, such as the study of convection in a spherical 
shell in mantle dynamics \citep{Ribe2018}. Indeed, symmetry 
arguments can be similarly exploited in a spherical geometry to understand the 
nature of the bifurcations \citep{Chossat1991,Matthews2003}. 

\subsection{Time dimension}

This work focuses on steady-states and thus there is little discussion of the 
time dimension. However, it is worth noting that while the governing equations 
are invariant under any time translation, they are not invariant under time 
reflection $m_t$ owing to the diffusion term.   

\subsection{Self-adjointness}

If the equations \eqref{eq:g1}, \eqref{eq:g2}
and \eqref{eq:g3} are linearised about a conductive steady-state (i.e. 
neglecting the $\v \bcdot \bnabla \theta$ term) then the equations themselves 
have an important symmetry: namely, they are self-adjoint provided the 
viscosity is constant or purely-depth dependent, and appropriate 
boundary conditions are applied.

\section{Representations of layer groups with non-zero 
wavevector}\label{sec:rep_nonzero}

The general theory of representations of layer groups with a non-zero 
wavevector is somewhat involved, and a full account can be found in e.g. 
\citet{Bradley1972, Aroyo2006b, DelaFlor2021, Grenier2012}. In this appendix we 
give some 
simple 
examples of representations with a non-zero 
wavenumber that are associated with spatial-period-multiplying bifurcations.

Consider the layer group $p4mm$ (No. 55). This group can be generated by unit 
translations $t_x$ and $t_y$ in the $x-$ and $y-$ directions along with point 
group element $4_z$ representing a 90 degree rotation about the $z$ axis and 
point group element 
$m_{\overline{x}y}$ representing a reflection in a vertical mirror plane 
parallel to the line $y=x$. A representation of the group can be described by 
a mapping of the generators to a set of matrices.

\begin{figure}
\centering
 \includegraphics[width=0.5\columnwidth]{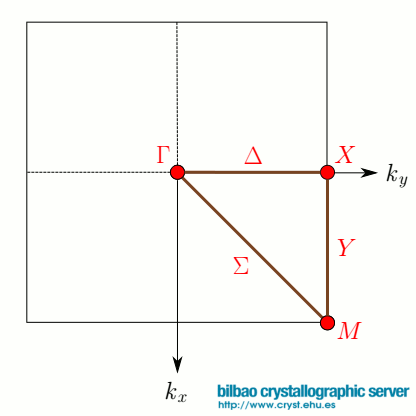}
 \caption{The Brillouin zone for $p4/nmm$ from the Bilbao Crystallographic 
Server \citep{DelaFlor2021}. The irreps are specified by a wavevector 
lying in the labelled triangular region, known as the representation domain. 
The origin is at the $\Gamma$ point. The special point $M$ is at 
$\left(\tfrac{1}{2}, \tfrac{1}{2}\right)$. Software which uses text labels will 
refer to $\Gamma$ as GM, $\Delta$ as DT, and $\Sigma$ as 
SM.}\label{fig:brillouin}
\end{figure}

In the general theory of representations of layer groups, each irreducible 
representation is described by a wavevector $\boldsymbol{k}$ and a label for a 
particular small 
representation of the wavevector. For this example we consider a wavevector of 
the form
\begin{equation}
 \boldsymbol{k} = u \boldsymbol{a}_1^* + u \boldsymbol{a}_2^* 
\label{eq:wave_vec}
\end{equation}
where $u<1/2$, and $\boldsymbol{a}_1^*$, $\boldsymbol{a}_2^*$ are basis vectors 
for 
the 
reciprocal lattice, defined such that $\boldsymbol{a}_i \bcdot 
\boldsymbol{a}_j^* = 2 
\upi \delta_{ij}$, where $\boldsymbol{a}_1$ and $\boldsymbol{a}_2$ are the 
space 
domain 
basis lattice vectors. The wavevector lies in a subset of the Brillouin 
zone known as the representation domain (\autoref{fig:brillouin}). 

The chosen wavevector in \eqref{eq:wave_vec} lies within the part of the 
Brillouin zone labelled 
$\Sigma$ (written in software which uses text labels as SM). The tool LKVEC 
\citep{DelaFlor2021} on the Bilbao Crystallographic Server can be used to 
identify the position of a wavenumber vector in the representation domain and 
the corresponding little co-group $\overline{G}^{\boldsymbol{k}}$. The little 
co-group is the set of point group 
elements that leaves the wavevector unchanged. The little
co-group associated with wavevectors along $\Sigma$ is $..m$, which has 
just two elements: the identity and the mirror 
$m_{\overline{x}y}$.  

From the little co-group $\overline{G}^{\boldsymbol{k}}$ one can form the 
little 
group $G^{\boldsymbol{k}}$ of 
$\boldsymbol{k}$, which a subgroup of $G$ containing those elements which have 
the 
point-group elements of the little co-group in their rotational part. We first 
need to obtain representations of the little group. Such representations must 
be small (or allowed) representations, which are the representations of the 
little group that map a pure 
translation by a vector $\boldsymbol{d}$ to $\exp(- i \boldsymbol{k} 
\bcdot \boldsymbol{d})$ 
times an identity matrix. In this simple example, the small representations are 
1-dimensional and there are just two of them. The small group is generated by 
the unit translations and the mirror element $m_{\overline{x}y}$. Since the 
representations of the translations has been prescribed, all that remains is to 
describe the mapping of the mirror element. There is a trivial representation 
$\Sigma_1$ which maps the mirror element $m_{\overline{x}y}$ to 1 and another 
representation $\Sigma_2$ which maps the mirror element to -1.

The \emph{star} of the wavevector is the set of possible wavevectors that can 
be obtained by applying all the point group operations to the given wavevector. 
In this example the star has four arms:
\begin{align}
  \boldsymbol{k}_1 &= (u, u), \\
  \boldsymbol{k}_2 &= (-u, -u), \\
  \boldsymbol{k}_3 &= (u, -u), \\
  \boldsymbol{k}_4 &= (-u, u).
\end{align}
The left cosets of $G^{\boldsymbol{k}}$ in $G$ are in one-to-one correspondence 
with the star of the wavevector. A representation of the full group $G$ can be 
obtained as an induced 
representation from the little group $G^{\boldsymbol{k}}$. The induced 
representation is of dimension $m d$ where $d$ is the dimension of the little 
group representation (here $d=1$) and $m$ is the number of left cosets of 
$G^{\boldsymbol{k}}$ 
in $G$ (which is identical to the number of arms in the star of the 
wavevector, here $m=4$). 

In the induced representation a general translation by a vector $\boldsymbol{d} 
= 
(d_1, d_2)$ is represented by a diagonal matrix of the following form
\begin{equation}
\begin{pmatrix}
 \e^{-\i \boldsymbol{k}_1 \bcdot \boldsymbol{d}} & 0 & 0 & 0 
\\
 0 & \e^{-\i \boldsymbol{k}_2 \bcdot  \boldsymbol{d}} & 0 & 0 \\
 0 & 0 & \e^{-\i \boldsymbol{k}_3 \bcdot  \boldsymbol{d}} & 0  \\
 0 & 0 & 0 & \e^{-\i \boldsymbol{k}_4 \bcdot  \boldsymbol{d}} \\
\end{pmatrix}
\end{equation}
where $\boldsymbol{k}_1$, $\boldsymbol{k}_2$, $\boldsymbol{k}_3$, and 
$\boldsymbol{k}_4$ are 
the 
arms of the star. 

The induced representation $^*\Gamma_1$ of the full space group $G$ is given in 
terms of the generators as
\begin{gather}
t_x = \begin{pmatrix}
 \omega & 0 & 0 & 0 \\
 0 & \omega^* & 0 & 0 \\
 0 & 0 & \omega & 0  \\
 0 & 0 & 0 & \omega^* \\
\end{pmatrix}, \quad
t_y = \begin{pmatrix}
 \omega & 0 & 0 & 0 \\
 0 & \omega^* & 0 & 0 \\
 0 & 0 & \omega^* & 0  \\
 0 & 0 & 0 & \omega \\
\end{pmatrix}, \nonumber \\
4_z = \begin{pmatrix}
 0 & 0 & 1 & 0 \\
 0 & 0 & 0 & 1 \\
 0 & 1 & 0 & 0  \\
 1 & 0 & 0 & 0 \\
\end{pmatrix}, \quad
m_{\overline{x}y} = \begin{pmatrix}
 1 & 0 & 0 & 0 \\
 0 & 1 & 0 & 0 \\
 0 & 0 & 0 & 1  \\
 0 & 0 & 1 & 0 \\
\end{pmatrix}, \label{eq:period_mult_rep}
\end{gather}
where $\omega = \e^{-2 \upi \i u}$. The representation $^*\Gamma_2$ has 
identical 
generators, except for 
$m_{\overline{x}y}$ which has $-1$ in place of $1$ in each entry. The above 
representation has complex entries, but by a change of basis an equivalent real 
representation can be found.

If the wavevector of form \eqref{eq:wave_vec} is chosen with $u=1/2$, then it 
lies 
at a special point in the Brillouin zone labelled $M$. The little co-group 
is then $4mm$, and the induced representation $^*M_1$ from the trivial 
representation of the little group is simply the 1-dimensional representation 
that maps the generators as
\begin{gather}
 t_x = -1, \quad t_y = -1, \quad 4_z = 1, \quad m_{\overline{x}y} = 1. 
\label{eq:period_mult_rep2}
\end{gather}
There are four additional representations induced from the little group, 
including a 2-dimensional representation $^*M_5$, but these will not be 
considered further.

\subsection{Application to spatial-period-multiplying bifurcations}

The representations $^*\Gamma_1$ and $^*M_1$ with matrices given in 
\eqref{eq:period_mult_rep} and \eqref{eq:period_mult_rep2} can be used to 
describe spatial-period-multiplying 
bifurcations for $p4mm$. The simplest case is when $u = 1/2$ and the 
1-D representation $^*M_1$ in \eqref{eq:period_mult_rep2} provides a mapping 
onto the group $C_2$. There is an 
isotropy subgroup which consists of all those elements which map to 1. This 
subgroup is also $p4mm$ (as all the point group elements are 
retained) but it has a reduced set of translation elements (it is an index 2 
klassengleiche subgroup). For the subgroup a 
new basis for the lattice can be obtained using the translations by $(1, 1)$ 
and 
$(1, -1)$. This isotropy subgroup is the maximal isotypic subgroup of lowest 
index for $p4mm$.

For the representation $^*\Gamma_1$, suppose $u=1/p$ where $p$ is a prime 
number equal to 3 
or greater. Then $\omega^p = 1$, and the matrix group described by 
\eqref{eq:period_mult_rep} is the finite group $D_4 \ltimes C_p^2$ of order $8 
p^2$. There is a klassengleiche axial isotropy subgroup of this representation 
of $p4mm$ which is also 
$p4mm$ but with the basis vectors of the lattice scaled by a factor of $p$ in 
each direction (the index of the subgroup in the parent group is $p^2$). All 
the point group operations are retained in the subgroup. In the representation 
in 
\eqref{eq:period_mult_rep} this can be explicitly recognised by the fixed point 
subspace $(a, a, a, a)$ which is invariant under $4_z$ and $m_{\overline{x}y}$ 
(retaining the point group) and the translations for which $(d_1, d_2) \equiv 
(0, 0) \,
(\operatorname{mod} p)$. There are also non-isotypic axial isotropy subgroups 
of $^*\Gamma_1$ with fixed point subspaces $(b, b, 0, 0)$ and $(0, 0, c, c)$. 
For these isotropy subgroups the point group is reduced to $mm2$, consisting of 
the diagonal mirrors and $2_z$. The translation group is reduced, but retains 
the diagonal translation with multiples of either $(1, -1)$ or $(1, 1)$. The 
corresponding layer group is $cmm2$ (index $2p$ subgroup).

For the particular case of $p=3$, the character table for the group $D_4 
\ltimes C_3^2$ is given in supplement 3 and further discussion of its role in 
spatial-period-multiplying bifurcations can be found in \citet{Matthews2004} 
(see their Figure 1).

\subsection{An example of $p4/nmm$}\label{sec:p4nmm_reps}

A slightly more complicated, but closely related, example is given by $p4/nmm$ 
(No. 64). This is non-symmorphic layer group. It can be generated by the same 
operations as $p4mm$, i.e. $t_x$, $t_y$, $4_z$, and $m_{\overline{x}y}$, but 
in addition is generated by a glide reflection $n$ which reflects in a vertical 
mirror $m_z$ and then translates by $\left(\tfrac{1}{2}, \tfrac{1}{2}, 
0\right)$. An index-9 
k-transition from $p4/nmm$ to $p4/nmm$ can be obtained as an isotropy subgroup 
of two different irreps. The irrep $^* \Sigma_1$ with wavevector 
$\boldsymbol{k} = (1/3, 1/3)$ given by
\begin{gather}
t_x = \begin{pmatrix}
 \omega & 0 & 0 & 0 \\
 0 & \omega^* & 0 & 0 \\
 0 & 0 & \omega & 0  \\
 0 & 0 & 0 & \omega^* \\
\end{pmatrix}, \quad
t_y = \begin{pmatrix}
 \omega & 0 & 0 & 0 \\
 0 & \omega^* & 0 & 0 \\
 0 & 0 & \omega^* & 0  \\
 0 & 0 & 0 & \omega \\
\end{pmatrix}, \nonumber \\
4_z = \begin{pmatrix}
 0 & 0 & 1 & 0 \\
 0 & 0 & 0 & 1 \\
 0 & 1 & 0 & 0  \\
 1 & 0 & 0 & 0 \\
\end{pmatrix}, \quad
m_{\overline{x}y} = \begin{pmatrix}
 1 & 0 & 0 & 0 \\
 0 & 1 & 0 & 0 \\
 0 & 0 & 0 & 1  \\
 0 & 0 & 1 & 0 \\
\end{pmatrix},
 \quad
n = \begin{pmatrix}
 \omega & 0 & 0 & 0 \\
 0 & \omega^* & 0 & 0 \\
 0 & 0 & 1 & 0  \\
 0 & 0 & 0 & 1 \\
\end{pmatrix}, \label{eq:period_mult_rep3}
\end{gather}
where $\omega = \e^{-2 \upi \i /3}$, and the irrep $^* \Delta_3$ with 
wavevector $\boldsymbol{k} = (0, 1/3)$ given by
\begin{gather}
t_x = \begin{pmatrix}
 1 & 0 & 0 & 0 \\
 0 & 1 & 0 & 0 \\
 0 & 0 & \omega & 0  \\
 0 & 0 & 0 & \omega^* \\
\end{pmatrix}, \quad
t_y = \begin{pmatrix}
 \omega & 0 & 0 & 0 \\
 0 & \omega^* & 0 & 0 \\
 0 & 0 & 1 & 0  \\
 0 & 0 & 0 & 1 \\
\end{pmatrix}, \nonumber \\
4_z = \begin{pmatrix}
 0 & 0 & 1 & 0 \\
 0 & 0 & 0 & 1 \\
 0 & 1 & 0 & 0  \\
 1 & 0 & 0 & 0 \\
\end{pmatrix}, \quad
m_{\overline{x}y} = \begin{pmatrix}
 0 & 0 & 1 & 0 \\
 0 & 0 & 0 & 1 \\
 1 & 0 & 0 & 0  \\
 0 & 1 & 0 & 0 \\
\end{pmatrix}, \quad
n = \begin{pmatrix}
 \omega^* & 0 & 0 & 0 \\
 0 & \omega & 0 & 0 \\
 0 & 0 & \omega^* & 0  \\
 0 & 0 & 0 & \omega \\
\end{pmatrix}. \label{eq:period_mult_rep4}
\end{gather}
For both representations \eqref{eq:period_mult_rep3} and 
\eqref{eq:period_mult_rep4} $(a, a, a, a)$ is the fixed point subspace 
corresponding to the axial isotropy subgroup $p4/nmm$. Note that for both cases 
$t_x^3$, $t_y^3$, and $n^3$ map to the identity and so are elements of the 
isotropy subgroup. $n^3$ represents a reflection in a vertical mirror followed 
by a translation by  $(3/2, 3/2, 0)$, so is the same as the original glide 
reflection $n$ but with the translation vector scaled by 3.

\section{Equivariants and character theory}\label{sec:equi}

Much of the key information about symmetry-breaking bifurcations can be obtained 
from a 
series of routine mechanical calculations using character tables. As described 
by \citet{Matthews2004} and \citet{Antoneli2008}, these calculations 
can be automated using the computational algebra package GAP \citep{gap2021}. 

Many key results follow from the \textit{trace formula} that states 
that for a group $G$ acting linearly on a vector space $V$ the dimension of the 
fixed-point subspace is
\begin{equation}
\text{dim}\; \text{Fix}(G, V) = \langle \chi_V, 1\rangle
\end{equation}
where $\chi_V$ is the character of the representation of $G$ on $V$, 1 is 
the trivial character, and angle brackets represent the scalar product on 
characters. This formula can be used to determine whether a subgroup is an 
isotropy subgroup \citep{Matthews2004}. Note that $V$ in the trace formula can 
be any vector space, 
not just $\mathbb{R}^n$. By applying this formula to appropriately symmetrised 
parts of tensor product spaces, \citet{Antoneli2008} show how this formula can 
be used to work out the dimension of the spaces of 
invariant and equivariant polynomials (their (3.9) and (3.10)). If $I(k)$ is the 
dimension of the space of invariant polynomials of degree $k$, and $E(k)$ is 
the corresponding space of equivariants of degree $k$, then the trace formula 
yields
\begin{gather}
 I(k) = \left\langle \chi_{S^k V}, 1 \right\rangle, \\
 E(k) = \left\langle \chi_{S^k V} \chi_{V}, 1 \right\rangle,
\end{gather}
where $S^k V$ refers to the symmetric part of the tensor product of $k$ copies 
of $V$. 

As an example, suppose we want to work out the number of quadratic equivariants for the faithful irrep of $D_3$. The character of the irrep can be written as $\chi_V = (2, 0, -1)$ where the identity corresponds to 2, the mirrors correspond to 0, and the rotations correspond to $-1$. The symmetric part of $V \otimes V$ has character $\chi_{S^2 V} = (3, 1, 0)$. Thus 
$\chi_{S^2 V} \chi_V = (6, 0, 0)$. The inner product with the trivial character 
then yields $E(2) = 6 / 6 = 1$ (since the order of $D_3$ is 6). Supplement 3 
provides tables of $I(k)$ and $E(k)$ for a series of small finite groups.

\end{document}


\title{Supplement 3: Character Tables}
\date{}
\maketitle

This supplement provides character tables for some of the small finite groups 
associated with symmetry-breaking 
bifurcations. The character tables were calculated using the GAP software with 
an algorithm for calculating isotropy subgroups described by 
\citet{Matthews2004}. 

The first table on each page is a standard character table but with an extra 
column on the 
right, following a similar format as those in \citet{Matthews2004}. The header 
of the table `1a', `2a', etc. lists the conjugacy classes where the number 
indicates the order of the elements. The next row (labelled `size') gives the 
number of elements in each conjugacy class. For some of the tables, the next 
rows give the power map, with rows labelled `2P', `3P', etc. These give the 
conjugacy class when a given element is raised to a certain prime-order power. 
The remaining rows give the characters of the irreps, labelled $R_1$, $R_2$, 
etc. A letter $F$ is added to denote that the representation is faithful.

The column on the far right 
gives the axial isotropy subgroups associated with each representation. Where 
numbers in brackets 
are given, the quotient group $N_\Gamma(\Sigma)/\Sigma$ is non-trivial and the 
number gives the order of the quotient group. Here $N_\Gamma(\Sigma)$ is the 
normalizer, defined by
\begin{equation}
 N_\Gamma(\Sigma) = \left\lbrace \gamma \in \Gamma : \gamma^{-1} \Sigma \gamma 
= \Sigma \right \rbrace,
\end{equation}
$\Gamma$ is the parent group and $\Sigma$ is the isotropy subgroup. Those 
isotopy subgroups with (2) denoted will be associated 
with pitchfork bifurcations. Those without a number may be transcritical, but 
having $N_\Gamma(\Sigma)/\Sigma = 1$ is not a sufficient condition for being 
transcritical. $D_5$ provides an example where the bifurcation associated with 
the faithful irrep is not transcritical, but $N_\Gamma(\Sigma)/\Sigma = 1$.

The second table gives the dimensions of the spaces of polynomial invariants 
$I(k)$ of degree $k$, and the third table gives the dimensions of the spaces of 
polynomial equivariants $E(k)$ of degree $k$. These tables help further 
classify the types of 
bifurcation and can be used to constrain the number of free parameters 
needed in the amplitude equations. 

\bibliography{groups}
\bibliographystyle{plainnat}

\clearpage
\section*{$C_2$}
\begin{table}[!ht]
\centering
\begin{tabular}{r|rr|l}
\toprule
 & 1a & 2a & axial subgroups \\
\midrule
size & 1 & 1 &  \\
\midrule$R_{1}$ & $1$ & $1$ & $C_2$ \\
$R_{2}$F & $1$ & $-1$ & $C_1$(2) \\
\bottomrule
\end{tabular}
\end{table}
\begin{table}[!ht]
\centering
\begin{tabular}{r|rrrrr}
\toprule
 & $I(2)$ & $I(3)$ & $I(4)$ & $I(5)$ & $I(6)$ \\
\midrule
$R_{1}$ & 1 & 1 & 1 & 1 & 1 \\
$R_{2}$F & 1 & 0 & 1 & 0 & 1 \\
\bottomrule
\end{tabular}
\end{table}
\begin{table}[!ht]
\centering
\begin{tabular}{r|rrrrr}
\toprule
 & $E(1)$ & $E(2)$ & $E(3)$ & $E(4)$ & $E(5)$ \\
\midrule
$R_{1}$ & 1 & 1 & 1 & 1 & 1 \\
$R_{2}$F & 1 & 0 & 1 & 0 & 1 \\
\bottomrule
\end{tabular}
\end{table}

The faithful irrep $R_2F$ is associated with a pitchfork bifurcation.
\clearpage

\section*{$D_3$}
\begin{table}[!ht]
\centering
\begin{tabular}{r|rrr|l}
\toprule
 & 1a & 2a & 3a & axial subgroups \\
\midrule
size & 1 & 3 & 2 &  \\
\midrule2P & 1a & 1a & 3a &  \\
3P & 1a & 2a & 1a &  \\
\midrule$R_{1}$ & $1$ & $1$ & $1$ & $D_3$ \\
$R_{2}$ & $1$ & $-1$ & $1$ & $C_3$(2) \\
$R_{3}$F & $2$ & $0$ & $-1$ & $C_2$ \\
\bottomrule
\end{tabular}
\end{table}
\begin{table}[!ht]
\centering
\begin{tabular}{r|rrrrr}
\toprule
 & $I(2)$ & $I(3)$ & $I(4)$ & $I(5)$ & $I(6)$ \\
\midrule
$R_{1}$ & 1 & 1 & 1 & 1 & 1 \\
$R_{2}$ & 1 & 0 & 1 & 0 & 1 \\
$R_{3}$F & 1 & 1 & 1 & 1 & 2 \\
\bottomrule
\end{tabular}
\end{table}
\begin{table}[!ht]
\centering
\begin{tabular}{r|rrrrr}
\toprule
 & $E(1)$ & $E(2)$ & $E(3)$ & $E(4)$ & $E(5)$ \\
\midrule
$R_{1}$ & 1 & 1 & 1 & 1 & 1 \\
$R_{2}$ & 1 & 0 & 1 & 0 & 1 \\
$R_{3}$F & 1 & 1 & 1 & 2 & 2 \\
\bottomrule
\end{tabular}
\end{table}

The faithful irrep $R_3F$ is associated with a transcritical bifurcation.
\clearpage
\section*{$D_4$}
\begin{table}[!ht]
\centering
\begin{tabular}{r|rrrrr|l}
\toprule
 & 1a & 2a & 2b & 2c & 4a & axial subgroups \\
\midrule
size & 1 & 1 & 2 & 2 & 2 &  \\
\midrule$R_{1}$ & $1$ & $1$ & $1$ & $1$ & $1$ & $D_4$ \\
$R_{2}$ & $1$ & $1$ & $-1$ & $-1$ & $1$ & $C_4$(2) \\
$R_{3}$ & $1$ & $1$ & $1$ & $-1$ & $-1$ & $C_2^2$(2) \\
$R_{4}$ & $1$ & $1$ & $-1$ & $1$ & $-1$ & $C_2^2$(2) \\
$R_{5}$F & $2$ & $-2$ & $0$ & $0$ & $0$ & $C_2$(2), $C_2$(2) \\
\bottomrule
\end{tabular}
\end{table}
\begin{table}[!ht]
\centering
\begin{tabular}{r|rrrrr}
\toprule
 & $I(2)$ & $I(3)$ & $I(4)$ & $I(5)$ & $I(6)$ \\
\midrule
$R_{1}$ & 1 & 1 & 1 & 1 & 1 \\
$R_{2}$ & 1 & 0 & 1 & 0 & 1 \\
$R_{3}$ & 1 & 0 & 1 & 0 & 1 \\
$R_{4}$ & 1 & 0 & 1 & 0 & 1 \\
$R_{5}$F & 1 & 0 & 2 & 0 & 2 \\
\bottomrule
\end{tabular}
\end{table}
\begin{table}[!ht]
\centering
\begin{tabular}{r|rrrrr}
\toprule
 & $E(1)$ & $E(2)$ & $E(3)$ & $E(4)$ & $E(5)$ \\
\midrule
$R_{1}$ & 1 & 1 & 1 & 1 & 1 \\
$R_{2}$ & 1 & 0 & 1 & 0 & 1 \\
$R_{3}$ & 1 & 0 & 1 & 0 & 1 \\
$R_{4}$ & 1 & 0 & 1 & 0 & 1 \\
$R_{5}$F & 1 & 0 & 2 & 0 & 3 \\
\bottomrule
\end{tabular}
\end{table}

The faithful irrep $R_5F$ is associated with a pitchfork bifurcation.

\clearpage
\section*{$D_5$}
\begin{table}[!ht]
\centering
\begin{tabular}{r|rrrr|l}
\toprule
 & 1a & 2a & 5a & 5b & axial subgroups \\
\midrule
size & 1 & 5 & 2 & 2 &  \\
\midrule2P & 1a & 1a & 5b & 5a &  \\
3P & 1a & 2a & 5b & 5a &  \\
5P & 1a & 2a & 1a & 1a &  \\
\midrule$R_{1}$ & $1$ & $1$ & $1$ & $1$ & $D_5$ \\
$R_{2}$ & $1$ & $-1$ & $1$ & $1$ & $C_5$(2) \\
$R_{3}$F & $2$ & $0$ & $\tfrac{-1-\sqrt{5}}{2}$ & $\tfrac{-1+\sqrt{5}}{2}$ & $C_2$ \\
$R_{4}$F & $2$ & $0$ & $\tfrac{-1+\sqrt{5}}{2}$ & $\tfrac{-1-\sqrt{5}}{2}$ & $C_2$ \\
\bottomrule
\end{tabular}
\end{table}
\begin{table}[!ht]
\centering
\begin{tabular}{r|rrrrr}
\toprule
 & $I(2)$ & $I(3)$ & $I(4)$ & $I(5)$ & $I(6)$ \\
\midrule
$R_{1}$ & 1 & 1 & 1 & 1 & 1 \\
$R_{2}$ & 1 & 0 & 1 & 0 & 1 \\
$R_{3}$F & 1 & 0 & 1 & 1 & 1 \\
$R_{4}$F & 1 & 0 & 1 & 1 & 1 \\
\bottomrule
\end{tabular}
\end{table}
\begin{table}[!ht]
\centering
\begin{tabular}{r|rrrrr}
\toprule
 & $E(1)$ & $E(2)$ & $E(3)$ & $E(4)$ & $E(5)$ \\
\midrule
$R_{1}$ & 1 & 1 & 1 & 1 & 1 \\
$R_{2}$ & 1 & 0 & 1 & 0 & 1 \\
$R_{3}$F & 1 & 0 & 1 & 1 & 1 \\
$R_{4}$F & 1 & 0 & 1 & 1 & 1 \\
\bottomrule
\end{tabular}
\end{table}

The faithful irreps $R_3F$ and $R_4F$ are associated with pitchfork 
bifurcations.

\clearpage
\section*{$S_4$}
\begin{table}[!ht]
\centering
\begin{tabular}{r|rrrrr|l}
\toprule
 & 1a & 2a & 2b & 3a & 4a & axial subgroups \\
\midrule
size & 1 & 3 & 6 & 8 & 6 &  \\
\midrule2P & 1a & 1a & 1a & 3a & 2a &  \\
3P & 1a & 2a & 2b & 1a & 4a &  \\
\midrule$R_{1}$ & $1$ & $1$ & $1$ & $1$ & $1$ & $S_4$ \\
$R_{2}$ & $1$ & $1$ & $-1$ & $1$ & $-1$ & $A_4$(2) \\
$R_{3}$ & $2$ & $2$ & $0$ & $-1$ & $0$ & $D_4$ \\
$R_{4}$F & $3$ & $-1$ & $-1$ & $0$ & $1$ & $C_2$(2), $C_3$(2), $C_4$(2) \\
$R_{5}$F & $3$ & $-1$ & $1$ & $0$ & $-1$ & $C_2^2$(2), $D_3$ \\
\bottomrule
\end{tabular}
\end{table}
\begin{table}[!ht]
\centering
\begin{tabular}{r|rrrrr}
\toprule
 & $I(2)$ & $I(3)$ & $I(4)$ & $I(5)$ & $I(6)$ \\
\midrule
$R_{1}$ & 1 & 1 & 1 & 1 & 1 \\
$R_{2}$ & 1 & 0 & 1 & 0 & 1 \\
$R_{3}$ & 1 & 1 & 1 & 1 & 2 \\
$R_{4}$F & 1 & 0 & 2 & 0 & 3 \\
$R_{5}$F & 1 & 1 & 2 & 1 & 3 \\
\bottomrule
\end{tabular}
\end{table}
\begin{table}[!ht]
\centering
\begin{tabular}{r|rrrrr}
\toprule
 & $E(1)$ & $E(2)$ & $E(3)$ & $E(4)$ & $E(5)$ \\
\midrule
$R_{1}$ & 1 & 1 & 1 & 1 & 1 \\
$R_{2}$ & 1 & 0 & 1 & 0 & 1 \\
$R_{3}$ & 1 & 1 & 1 & 2 & 2 \\
$R_{4}$F & 1 & 0 & 2 & 1 & 4 \\
$R_{5}$F & 1 & 1 & 2 & 2 & 4 \\
\bottomrule
\end{tabular}
\end{table}

The faithful irreps $R_4F$ is associated with a pitchfork bifurcation, the 
faithful irrep $R_5$ with a transcritical bifurcation.

\clearpage
\section*{$A_4$}
\begin{table}[!ht]
\centering
\begin{tabular}{r|rrrr|l}
\toprule
 & 1a & 2a & 3a & 3b & axial subgroups \\
\midrule
size & 1 & 3 & 4 & 4 &  \\
\midrule2P & 1a & 1a & 3b & 3a &  \\
3P & 1a & 2a & 1a & 1a &  \\
\midrule$R_{1}$ & $1$ & $1$ & $1$ & $1$ & $A_4$ \\
$R_{2}$ & $1$ & $1$ & $e^{4 \pi i/3}$ & $e^{2 \pi i/3}$ & $C_2^2$(3) \\
$R_{3}$ & $1$ & $1$ & $e^{2 \pi i/3}$ & $e^{4 \pi i/3}$ & $C_2^2$(3) \\
$R_{4}$F & $3$ & $-1$ & $0$ & $0$ & $C_2$(2), $C_3$ \\
\bottomrule
\end{tabular}
\end{table}
\begin{table}[!ht]
\centering
\begin{tabular}{r|rrrrr}
\toprule
 & $I(2)$ & $I(3)$ & $I(4)$ & $I(5)$ & $I(6)$ \\
\midrule
$R_{1}$ & 1 & 1 & 1 & 1 & 1 \\
$R_{2}$ & 0 & 1 & 0 & 0 & 1 \\
$R_{3}$ & 0 & 1 & 0 & 0 & 1 \\
$R_{4}$F & 1 & 1 & 2 & 1 & 4 \\
\bottomrule
\end{tabular}
\end{table}
\begin{table}[!ht]
\centering
\begin{tabular}{r|rrrrr}
\toprule
 & $E(1)$ & $E(2)$ & $E(3)$ & $E(4)$ & $E(5)$ \\
\midrule
$R_{1}$ & 1 & 1 & 1 & 1 & 1 \\
$R_{2}$ & 0 & 1 & 0 & 0 & 1 \\
$R_{3}$ & 0 & 1 & 0 & 0 & 1 \\
$R_{4}$F & 1 & 1 & 3 & 3 & 6 \\
\bottomrule
\end{tabular}
\end{table}

The faithful irrep $R_4F$ is associated with a transcritical bifurcation.

\clearpage
\section*{$C_5 \rtimes C_4$}
\begin{table}[!ht]
\centering
\begin{tabular}{r|rrrrr|l}
\toprule
 & 1a & 2a & 4a & 4b & 5a & axial subgroups \\
\midrule
size & 1 & 5 & 5 & 5 & 4 &  \\
\midrule2P & 1a & 1a & 2a & 2a & 5a &  \\
3P & 1a & 2a & 4b & 4a & 5a &  \\
5P & 1a & 2a & 4a & 4b & 1a &  \\
\midrule$R_{1}$ & $1$ & $1$ & $1$ & $1$ & $1$ & $C_5 \rtimes C_4$ \\
$R_{2}$ & $1$ & $1$ & $-1$ & $-1$ & $1$ & $D_5$(2) \\
$R_{3}$ & $1$ & $-1$ & $-i$ & $i$ & $1$ & $C_5$(4) \\
$R_{4}$ & $1$ & $-1$ & $i$ & $-i$ & $1$ & $C_5$(4) \\
$R_{5}$F & $4$ & $0$ & $0$ & $0$ & $-1$ & $C_4$ \\
\bottomrule
\end{tabular}
\end{table}
\begin{table}[!ht]
\centering
\begin{tabular}{r|rrrrr}
\toprule
 & $I(2)$ & $I(3)$ & $I(4)$ & $I(5)$ & $I(6)$ \\
\midrule
$R_{1}$ & 1 & 1 & 1 & 1 & 1 \\
$R_{2}$ & 1 & 0 & 1 & 0 & 1 \\
$R_{3}$ & 0 & 0 & 1 & 0 & 0 \\
$R_{4}$ & 0 & 0 & 1 & 0 & 0 \\
$R_{5}$F & 1 & 1 & 3 & 3 & 5 \\
\bottomrule
\end{tabular}
\end{table}
\begin{table}[!ht]
\centering
\begin{tabular}{r|rrrrr}
\toprule
 & $E(1)$ & $E(2)$ & $E(3)$ & $E(4)$ & $E(5)$ \\
\midrule
$R_{1}$ & 1 & 1 & 1 & 1 & 1 \\
$R_{2}$ & 1 & 0 & 1 & 0 & 1 \\
$R_{3}$ & 0 & 0 & 1 & 0 & 0 \\
$R_{4}$ & 0 & 0 & 1 & 0 & 0 \\
$R_{5}$F & 1 & 2 & 4 & 7 & 11 \\
\bottomrule
\end{tabular}
\end{table}

The faithful irrep $R_5F$ is associated with a transcritical bifurcation.

\clearpage
\section*{$C_3^2 \rtimes C_4$}
\begin{table}[!ht]
\centering
\begin{tabular}{r|rrrrrr|l}
\toprule
 & 1a & 2a & 3a & 3b & 4a & 4b & axial subgroups \\
\midrule
size & 1 & 9 & 4 & 4 & 9 & 9 &  \\
\midrule2P & 1a & 1a & 3a & 3b & 2a & 2a &  \\
3P & 1a & 2a & 1a & 1a & 4b & 4a &  \\
\midrule$R_{1}$ & $1$ & $1$ & $1$ & $1$ & $1$ & $1$ & $C_3^2 \rtimes C_4$ \\
$R_{2}$ & $1$ & $1$ & $1$ & $1$ & $-1$ & $-1$ & $C_3^2 \rtimes C_2$(2) \\
$R_{3}$ & $1$ & $-1$ & $1$ & $1$ & $-i$ & $i$ & $C_3 \times C_3$(4) \\
$R_{4}$ & $1$ & $-1$ & $1$ & $1$ & $i$ & $-i$ & $C_3 \times C_3$(4) \\
$R_{5}$F & $4$ & $0$ & $1$ & $-2$ & $0$ & $0$ & $C_4$, $D_3$ \\
$R_{6}$F & $4$ & $0$ & $-2$ & $1$ & $0$ & $0$ & $C_4$, $D_3$ \\
\bottomrule
\end{tabular}
\end{table}
\begin{table}[!ht]
\centering
\begin{tabular}{r|rrrrr}
\toprule
 & $I(2)$ & $I(3)$ & $I(4)$ & $I(5)$ & $I(6)$ \\
\midrule
$R_{1}$ & 1 & 1 & 1 & 1 & 1 \\
$R_{2}$ & 1 & 0 & 1 & 0 & 1 \\
$R_{3}$ & 0 & 0 & 1 & 0 & 0 \\
$R_{4}$ & 0 & 0 & 1 & 0 & 0 \\
$R_{5}$F & 1 & 1 & 2 & 2 & 4 \\
$R_{6}$F & 1 & 1 & 2 & 2 & 4 \\
\bottomrule
\end{tabular}
\end{table}
\begin{table}[!ht]
\centering
\begin{tabular}{r|rrrrr}
\toprule
 & $E(1)$ & $E(2)$ & $E(3)$ & $E(4)$ & $E(5)$ \\
\midrule
$R_{1}$ & 1 & 1 & 1 & 1 & 1 \\
$R_{2}$ & 1 & 0 & 1 & 0 & 1 \\
$R_{3}$ & 0 & 0 & 1 & 0 & 0 \\
$R_{4}$ & 0 & 0 & 1 & 0 & 0 \\
$R_{5}$F & 1 & 1 & 2 & 5 & 6 \\
$R_{6}$F & 1 & 1 & 2 & 5 & 6 \\
\bottomrule
\end{tabular}
\end{table}

The faithful irreps $R_5F$ and $R_6F$ are both associated with transcritical 
bifurcations.

\clearpage
\section*{$C_3^2 \rtimes D_4$}
\begin{table}[!ht]
\centering
\begin{tabular}{r|rrrrrrrrr|l}
\toprule
 & 1a & 2a & 2b & 2c & 3a & 3b & 4a & 6a & 6b & axial subgroups \\
\midrule
size & 1 & 6 & 6 & 9 & 4 & 4 & 18 & 12 & 12 &  \\
\midrule2P & 1a & 1a & 1a & 1a & 3a & 3b & 2c & 3a & 3b &  \\
3P & 1a & 2a & 2b & 2c & 1a & 1a & 4a & 2a & 2b &  \\
5P & 1a & 2a & 2b & 2c & 3a & 3b & 4a & 6a & 6b &  \\
\midrule$R_{1}$ & $1$ & $1$ & $1$ & $1$ & $1$ & $1$ & $1$ & $1$ & $1$ & $C_3^2 \rtimes D_4$ \\
$R_{2}$ & $1$ & $-1$ & $-1$ & $1$ & $1$ & $1$ & $1$ & $-1$ & $-1$ & $C_3^2 \rtimes C_4$(2) \\
$R_{3}$ & $1$ & $-1$ & $1$ & $1$ & $1$ & $1$ & $-1$ & $-1$ & $1$ & $D_3^2$(2) \\
$R_{4}$ & $1$ & $1$ & $-1$ & $1$ & $1$ & $1$ & $-1$ & $1$ & $-1$ & $D_3^2$(2) \\
$R_{5}$ & $2$ & $0$ & $0$ & $-2$ & $2$ & $2$ & $0$ & $0$ & $0$ & $C_3 \times D_3$(2), $C_3 \times D_3$(2) \\
$R_{6}$F & $4$ & $-2$ & $0$ & $0$ & $1$ & $-2$ & $0$ & $1$ & $0$ & $C_2^2$(2), $C_4$(2), $D_3$(2), $C_6$(2) \\
$R_{7}$F & $4$ & $0$ & $-2$ & $0$ & $-2$ & $1$ & $0$ & $0$ & $1$ & $C_2^2$(2), $C_4$(2), $D_3$(2), $C_6$(2) \\
$R_{8}$F & $4$ & $0$ & $2$ & $0$ & $-2$ & $1$ & $0$ & $0$ & $-1$ & $D_4$, $D_6$ \\
$R_{9}$F & $4$ & $2$ & $0$ & $0$ & $1$ & $-2$ & $0$ & $-1$ & $0$ & $D_4$, $D_6$ \\
\bottomrule
\end{tabular}
\end{table}
\begin{table}[!ht]
\centering
\begin{tabular}{r|rrrrr}
\toprule
 & $I(2)$ & $I(3)$ & $I(4)$ & $I(5)$ & $I(6)$ \\
\midrule
$R_{1}$ & 1 & 1 & 1 & 1 & 1 \\
$R_{2}$ & 1 & 0 & 1 & 0 & 1 \\
$R_{3}$ & 1 & 0 & 1 & 0 & 1 \\
$R_{4}$ & 1 & 0 & 1 & 0 & 1 \\
$R_{5}$ & 1 & 0 & 2 & 0 & 2 \\
$R_{6}$F & 1 & 0 & 2 & 0 & 4 \\
$R_{7}$F & 1 & 0 & 2 & 0 & 4 \\
$R_{8}$F & 1 & 1 & 2 & 2 & 4 \\
$R_{9}$F & 1 & 1 & 2 & 2 & 4 \\
\bottomrule
\end{tabular}
\end{table}
\begin{table}[!ht]
\centering
\begin{tabular}{r|rrrrr}
\toprule
 & $E(1)$ & $E(2)$ & $E(3)$ & $E(4)$ & $E(5)$ \\
\midrule
$R_{1}$ & 1 & 1 & 1 & 1 & 1 \\
$R_{2}$ & 1 & 0 & 1 & 0 & 1 \\
$R_{3}$ & 1 & 0 & 1 & 0 & 1 \\
$R_{4}$ & 1 & 0 & 1 & 0 & 1 \\
$R_{5}$ & 1 & 0 & 2 & 0 & 3 \\
$R_{6}$F & 1 & 0 & 2 & 1 & 5 \\
$R_{7}$F & 1 & 0 & 2 & 1 & 5 \\
$R_{8}$F & 1 & 1 & 2 & 4 & 5 \\
$R_{9}$F & 1 & 1 & 2 & 4 & 5 \\
\bottomrule
\end{tabular}
\end{table}

The faithful irreps $R_6F$ and $R_7F$ are associated with pitchfork 
bifurcations, the irreps $R_7F$ and $R_8F$ with transcritical bifurcations.

